\documentclass[namedreferences]{solarphysics}
\usepackage[optionalrh,solaenum]{spr-sola-addons} % For Solar Physics 
\pdfoutput=1
\usepackage{graphicx}        % For eps figures, newer & more powerfull
\usepackage[usenames]{color}           % For color text: \color command
\usepackage{url}             % For breaking URLs easily trough lines
\renewcommand{\vec}[1]{{\mathbfit #1}}
\newcommand{\deriv}[2]{\frac{{\mathrm d} #1}{{\mathrm d} #2}}
\newcommand{\rmd}{ {\ \mathrm d} }
\newcommand{\uvec}[1]{ \hat{\mathbf #1} }

\newcommand{\grad}{ {\bf \nabla } }
\newcommand{\curl}{ {\bf \nabla} \times}
\newcommand{\vol}{ {\mathcal V} }
\newcommand{\surf}{ {\mathcal S} }

\newcommand{\dv}{~{\mathrm d}^3 x}

\newcommand{\intv}{\int_{\vol}^{}}

\newcommand{\avec}{ \vec A}

\newcommand{\bb}{\vec B}

\newcommand{\xx}{ \vec x}
\newcommand{\uu}{ \vec u}

\newcommand{\BE}{\begin{equation}}
\newcommand{\EE}{\end{equation}}
\newcommand{\BA}{\begin{eqnarray}}
\newcommand{\EA}{\end{eqnarray}}
\newcommand{\vA}{\vec{A}}

\newcommand{\vAp}{\vA_{\rm p}}

\newcommand{\vB}{\vec{B}}

\newcommand{\vBp}{\vB_{\rm p}}
\newcommand{\Bn}{B_{n}}

\newcommand{\ga}{G_A}
\newcommand{\gth}{G_{\theta}}
\newcommand{\gph}{G_{\Phi}}
\newcommand{\Hm}{ {\mathcal L} }
\newcommand{\LLa}{{\mathcal L}^{\rm arch}}

\newcommand{\eq}[1]{Equation~(\ref{eq:#1})} 
\newcommand{\eqs}[2]{Equations~(\ref{eq:#1}) and (\ref{eq:#2})} 
\newcommand{\eqss}[2]{Equations~(\ref{eq:#1}) -- (\ref{eq:#2})} 
\newcommand{\sect}[1]{Section~\ref{sec:#1}} 
\newcommand{\sects}[2]{Sections~\ref{sec:#1} and \ref{sec:#2}} 
 
\newcommand{\app}[1]{Appendix~\ref{app:#1}}

\newcommand{\fig}[1]{Figure~\ref{fig:#1}}
 
\newcommand{\figpnls}[3]{Figures~\ref{fig:#1}{#2} and \ref{fig:#1}{#3}} 
\newcommand{\figss}[2]{Figures~\ref{fig:#1} -- \ref{fig:#2}} 
\newcommand{\cf}{\textit{cf.} }
\newcommand{\eg}{\textit{e.g.}, }
\newcommand{\ie}{\textit{i.e.}, }
\newcommand{\resp}{resp. }

\begin{document}

\begin{article}

\begin{opening}

\title{Photospheric Injection of Magnetic Helicity: Connectivity--based Flux Density Method}

\author{K.~\surname{Dalmasse}\sep
        E.~\surname{Pariat}\sep
        P.~\surname{D\'emoulin}\sep  
        G.~\surname{Aulanier}\sep            
       }
\runningauthor{Dalmasse et al.}
\runningtitle{connectivity-based magnetic helicity flux density}

   \institute{LESIA, Observatoire de Paris, CNRS, UPMC, Universit\'e Paris-Diderot, 92195 Meudon, France
                     email: \url{kevin.dalmasse@obspm.fr} \\ 
             }

	\begin{abstract}
Magnetic helicity quantifies how globally sheared and/or twisted is the magnetic field in a volume. 
This quantity is believed to play a key role in solar activity due to its conservation property. 
Helicity is continuously injected into the corona during the evolution 
of active regions (ARs). To better understand and quantify the role of 
magnetic helicity in solar activity, the distribution 
of magnetic helicity flux in ARs needs to be studied. 
The helicity distribution can be computed from the temporal evolution of photospheric 
magnetograms of ARs such as the ones provided by SDO/HMI and Hinode/SOT. Most 
recent analyses of photospheric helicity flux derive an helicity flux density proxy based 
on the relative rotation rate of photospheric magnetic footpoints. 
Although this proxy allows a good estimate of the photospheric helicity flux, it is still not 
a true helicity flux density because it does not take into account the connectivity of the 
magnetic field lines.  For the first time, we implement a helicity density which takes 
into account such connectivity. 
In order to use it for future observational studies, we test the method and its precision on 
several types of models involving different patterns of helicity injection. 
We also test it on more complex configurations --- from magnetohydrodynamics (MHD) 
simulations --- containing quasi--separatrix layers. 
We demonstrate that this connectivity--based helicity flux density proxy is the best to 
map the true distribution of photospheric helicity injection.
	\end{abstract}
\end{opening}
%-------------------------------------------------

\section{Introduction} \label{sec:Intro}

   %{\S}{\bf --- Introduction of H.  Interest?} \\

Magnetic helicity plays a key role in solar MHD 
because it is quasi conserved on timescales much smaller than the global
energy diffusion timescale (\opencite{Berger84}, \citeyear{Berger03}).  
This conservation property constrains the evolution of the magnetic field.
In particular, an isolated magnetic field structure with a non-null helicity cannot relax to 
a potential field, even through resistive mechanisms: its minimum energy is 
theoretically and experimentally bounded by a 
linear force-free field rather than the potential one \cite{Taylor74,Yamada99}. 
Linton and Antiochos (\citeyear{Linton02,Linton05}) have shown that helicity 
can be used to predict which type of interaction can occur between 
reconnected flux tubes.

In order for a system to reach the lowest possible energy state, its helicity 
must be eventually carried away or annihilated. In the solar corona, important 
helicity carriers are the twisted magnetic structures associated with coronal 
mass ejections (CMEs) and magnetic clouds. \inlinecite{Rust94} and 
\inlinecite{Low97} have therefore hypothesized that CMEs could be the 
result of the global conservation of helicity within the solar atmosphere 
(see also \opencite{Zhang06}, \citeyear{Zhang12}; \opencite{Zhang08}). 
CMEs can transport the large amount of helicity present in 
ARs that have been injected from the solar interior 
(\opencite{vanDriel99}; \opencite{DeVore00a}; \opencite{Demoulin02a}; 
\opencite{Green02a}, \citeyear{Green02b}; \opencite{Mandrini04}; 
\opencite{Georgoulis09}; \opencite{Kazachenko12}).

   %{\S}{\bf --- Helicity annihilation} \\

Another possible way for a magnetic system to get its helicity 
content reduced is through annihilation by magnetic reconnection with other systems 
containing helicity of a different sign. Through these reconnections, the global system 
would present a lower absolute amount of magnetic helicity and thus the minimum
energy state that could be reached would be also lower. It has thus been conjectured 
that reconnection between systems of opposite magnetic helicity would lead to a higher 
energy release \cite{Kusano95}. \inlinecite{Linton01} have shown that reconnection 
between two twisted flux tubes is more violent and that more energy is released when 
the flux tubes have opposite helicity rather than the same sign. 
Models of flares involving opposite sign of helicity have thus been developed 
(\opencite{Kusano02}, \citeyear{Kusano04}) and observational studies aiming at detecting 
ARs with opposite helicity signs have been carried over 
\cite{Chandra10,Romano11a,Romano11b}.

   %{\S}{\bf --- Helicity obs gen} \\

Estimation of the magnetic helicity in the solar atmosphere is not straightforward 
(see reviews by \opencite{Demoulin07}; \opencite{Demoulin09}). The sign of 
magnetic helicity can be derived from the observation of twisted or sheared structures 
such as filaments fibrils and barbs, sunspot whorls, magnetic tongues.

   %{\S}{\bf --- Helicity flux} \\
   
A first way to quantitatively determine magnetic helicity is based on magnetic field 
extrapolations (\eg \opencite{Berger03}; \opencite{Demoulin07}; 
\opencite{Valori12}; \opencite{Jing12}). 
Another method to estimate the magnetic helicity has relied on the 
measurements of its flux through the solar surface \cite{Chae01,Chae04}. 
Magnetic helicity fluxes  can directly be estimated using sequences of longitudinal 
magnetograms \cite{Demoulin03}, with improved measurements obtained 
when full vector magnetograms at high enough cadence are available 
\cite{Schuck08,Yang12}. The estimation of the helicity flux has allowed 
to track the evolution of the helicity injected in many ARs 
(\eg \opencite{Nindos03}; \opencite{Yamamoto05}; \opencite{Jeong07}; 
\opencite{Labonte07}, \opencite{Yang12}).  

   %{\S}{\bf --- Helicity flux density} \\
In all studied ARs, an extremely mixed pattern of helicity injection is observed (\eg 
\opencite{Chae04}; \opencite{Yamamoto05}). However \inlinecite{Pariat05} showed 
that the analyzed quantity, $\ga$, was not a proper helicity flux density 
and that it produced important spurious non-physical signals. They showed that 
helicity flux density is inherently not a local quantity per unit 
surface. The physically meaningful helicity flux density 
is the helicity per elementary flux tube. They proposed two quantities 
named $\gth$ and $\gph$ (see their derivation in \sect{S-MagHFDensity}) 
that can be used as proxies of the helicity flux density through the photosphere.

     %{\S}{\bf --- Gtheta} \\
The first proxy, $\gth$, could be directly applied from time sequence of magnetograms 
(see \opencite{Chae07}; \opencite{Yang12}; for improved high--efficient methods to compute $\gth$). 
This proxy, removing the spurious signal of $\ga$, showed that the helicity flux 
injection pattern in AR was rather uniform in sign \cite{Pariat06}. Most ARs do not 
present important traces of injection of helicity of opposite sign. However, $\gth$ is 
not completely free of spurious signal (\opencite{Pariat05}, \citeyear{Pariat06}, 
\citeyear{Pariat07c}) and direct interpretations of the maps should be taken with caution.

     %{\S}{\bf --- Gphi} \\
The second proxy, $\gph$, is the proxy that allows the more truthful 
representation of the helicity flux distribution. It however requires the knowledge of the 
magnetic field connectivity. It has so far never been directly used with any observed data. 
Its implementation is indeed not straightforward in non-analytical fields as it requires a good 
knowledge of the 3D magnetic field. As 3D magnetic field extrapolations would become 
more common and hopefully more reliable, it will be more easy to use $\gph$ on 
observational cases.

 %{\S}{\bf --- Goal} \\
The present study is a first step toward implementing a method that would be used with 
observed data (\ie magnetic field extrapolations). 
The aim is to practically test the $\gph$ method on 
simplified solar configurations in order to determine typical patterns of $\gph$. 
This would help to later interpret observed $\gph$ maps. It will also help us to 
understand the limit and precision of the method. 
 
%{\S}{\bf --- Outline} \\
The manuscript is organized as follows: \sect{S-PhotHelFlux} presents the analytical derivation 
of $\gth$ and $\gph$ and summarizes some of the expected properties. In \sect{S-methodo}, we 
present the method and introduce the different magnetic configurations 
and flow patterns studied. Next, we present the results of our analyzes on different models:
first on two analytical configurations (\sects{S-two-source-charges}{S-Torus}), 
then on magnetic extrapolations (\sect{S-Uniform}), and finally on MHD simulations with 
complex connectivities (\sect{S-Hinj-in-QSLs}). We conclude in \sect{S-Conclusion}.

\section{Photospheric Helicity Flux} \label{sec:S-PhotHelFlux}      

\subsection{Magnetic Helicity Flux} \label{sec:S-MagHelFlux}
  
   %{\S}{\bf --- Hr def.} \\
   
Let $\vol $ be a magnetic volume bounded by a surface $\surf $, with 
magnetic flux crossing $\surf $ (\eg $\vol $ is part of the corona). A gauge 
invariant relative magnetic helicity, $H$, can be 
written as follows (\opencite{Finn85}):
    \BE   \label{eq:Eq-Hr-def}
     H = \intv (\avec + \vAp) \cdot (\bb - \vBp) \dv \,,
    \EE
where $\avec$ is the vector potential ($\bb = \curl \avec$). 
In this formula, the magnetic helicity is defined relatively to the potential magnetic 
field, $\vBp$ ($\vBp = \curl \vAp$), that has the same normal component ($\Bn$) 
on $\surf $ as $\bb$.

   %{\S}{\bf --- H Flux} \\
   
\inlinecite{Pariat05} demonstrated that the magnetic helicity flux across $\surf $ 
could be written as the summation of the relative rotation rate, on $\surf $, of all 
pairs of elementary magnetic flux tubes weighted by their magnetic flux:
    \BE \label{eq:Eq-Hthetaflux}
     \deriv{H}{t} = - \frac{1}{2\pi} \int_{\surf} \int_{\surf^\prime} 
	                         \deriv{\theta (\xx-\xx^\prime)}{t} \Bn (\xx) 
	                         \Bn (\xx^\prime) \rmd \surf \rmd \surf^\prime \,,
    \EE
where
    \BE
     \deriv{\theta (\xx-\xx^\prime)}{t} =\frac{ \left( (\xx-\xx') \times (\uu-\uu') \right) |_{n}}{|\xx-\xx'|^{2}}
    \EE
is the relative rotation rate between the two photospheric points $\xx$ and 
$\xx'$ moving on the photosphere with the flux transport velocity $\uu$ and 
$\uu'$ respectively.	

   %{\S}{\bf --- Estimating dH/dt from observations} \\ 

Observationally, a time series of magnetograms provides $\Bn$ at the photosphere. 
Several methods have been developed to estimate $\uu$. One is based on tracking 
the photospheric spatial evolution of magnetic flux tubes from magnetograms and is 
called Local Correlation Tracking (LCT; \inlinecite{Chae01} and references 
therein). Others are based on solving the induction equation using magnetograms 
\cite{Longcope04}. There are also methods that solve the induction equation 
in the spirit of the LCT method (\opencite{Welsch04}, 
\citeyear{Welsch07}; \opencite{Schuck05}, \citeyear{Schuck06,Schuck08}).

\subsection{Magnetic Helicity Flux Density} \label{sec:S-MagHFDensity}

   %{\S}{\bf --- Gtheta} \\ 
   
From \eq{Eq-Hthetaflux}, \inlinecite{Pariat05} defined a new helicity flux density 
proxy, $\gth$, that represents the distribution of helicity density at the 
photosphere:
    \BE \label{eq:Eq-Gtheta}
      \gth(\xx) = - \frac{\Bn}{2\pi} \int_{\surf^\prime} 
                                         \deriv{\theta (\xx-\xx^\prime)}{t} 
                                        \Bn^\prime \rmd \surf^\prime \,.
    \EE	
	
However, magnetic helicity is a global quantity. The helicity density and the density 
of helicity flux are only meaningful when considering a whole 
magnetic flux tube, which requires the knowledge of the magnetic connectivity 
in the volume $\vol $ (\opencite{Pariat05}).

  %{\S}{\bf --- Introducing a true helicity density flux} \\ 
	
Separating \eq{Eq-Hthetaflux} into two terms, \ie the flux of helicity due to the 
relative rotation of positive and negative polarities --- first term of \eq{Eq-HthF-bis} 
--- and the one due to the relative rotation of each polarity --- second term of 
\eq{Eq-HthF-bis}, they rewrite \eq{Eq-Hthetaflux} as follows:
     \BA \label{eq:Eq-HthF-bis}
     \deriv{H}{t} \quad &=& \quad \frac{1}{2\pi} \int \int_{\Bn \cdot \Bn^\prime < 0} 
	                         \deriv{\theta}{t} | \Bn \Bn^\prime | \rmd \surf \rmd \surf^\prime
	                         \nonumber \\
	                         && \:
	                        - \frac{1}{2\pi} \int \int_{\Bn \cdot \Bn^\prime > 0} 
	                         \deriv{\theta}{t} \Bn \Bn^\prime \rmd \surf \rmd \surf^\prime  \,.
      \EA
Using $\rmd \Phi_{+}=\Bn(\xx_{+}) \rmd \surf$ and $\rmd \Phi_{-}=- \Bn(\xx_{-}) \rmd \surf$ 
the elementary magnetic fluxes in the positive and negative polarity respectively, 
\eq{Eq-HthF-bis} leads to:
      \BA \label{eq:Eq-dHthF-ter}
      \deriv{H}{t} \quad &=& \quad \frac{1}{2\pi} \int_{\Phi_{+}} \int_{\Phi_{-}^\prime} 
	                         \deriv{\theta (\xx_{+}-\xx_{-}^\prime)}{t} \rmd \Phi_{+} \rmd \Phi_{-}^\prime
	                         \nonumber \\
	                         && \:
	                         + \frac{1}{2\pi} \int_{\Phi_{-}} \int_{\Phi_{+}^\prime} 
	                         \deriv{\theta (\xx_{-}-\xx_{+}^\prime)}{t} \rmd \Phi_{-} \rmd \Phi_{+}^\prime
	                         \nonumber \\
	                         && \:
	                        - \frac{1}{2\pi} \int_{\Phi_{+}} \int_{\Phi_{+}^\prime} 
	                         \deriv{\theta (\xx_{+}-\xx_{+}^\prime)}{t} \rmd \Phi_{+} \rmd \Phi_{+}^\prime
	                         \nonumber \\
	                         && \:
	                        - \frac{1}{2\pi} \int_{\Phi_{-}} \int_{\Phi_{-}^\prime} 
	                         \deriv{\theta (\xx_{-}-\xx_{-}^\prime)}{t} \rmd \Phi_{-} \rmd \Phi_{-}^\prime  \,.
      \EA

Since the magnetic flux is conserved along the flux tubes, we have $\rmd \Phi_{+}=\rmd \Phi_{-}$ 
and $\rmd \Phi_{+}^\prime=\rmd \Phi_{-}^\prime$. Then, by considering 
two generic magnetic field lines {\it ``a''} and {\it ``c''} respectively going from footpoint 
$\xx_{a_{+}}$ to $\xx_{a_{-}}$ and from $\xx_{c_{+}}$ to $\xx_{c_{-}}$ 
(\fig{Fig-Methodo-connecti}a), and by regrouping all four terms of \eq{Eq-dHthF-ter}, we 
can rewrite the helicity flux by explicitly including the field lines connectivity 
in $\vol $:
      \BA        \label{eq:Eq-dH-Phi-doubleInt}
      \deriv{H}{t} 
                    &=& \int_{\Phi} \deriv{h_{\Phi}}{t}\bigg|_{c} \rmd \Phi_{c_{+}} \,,
                         \label{eq:Eq-dH-Phi} \\
                         && \nonumber \\
                         && \nonumber \\
        \textrm{with }  \deriv{h_{\Phi}}{t}\bigg|_{c} &=& 
                    \frac{1}{2 \pi} \int_{\Phi}
                    \left(   \deriv{ \theta (\xx _{c_+} -\xx _{a_-}) }{t}
                           + \deriv{ \theta (\xx _{c_-} -\xx _{a_+}) }{t}
                    \right.                            \nonumber  \\
       && \quad \quad \quad \;
                    \left. - \deriv{ \theta (\xx _{c_+} -\xx _{a_+}) }{t}
                         - \deriv{ \theta (\xx _{c_-} -\xx _{a_-}) }{t}
                    \right) 
                    \rmd \Phi_{a_{+}} \,.
                     \label{eq:Eq-dhdens-Phi} 
       \EA 
              
In \eq{Eq-dH-Phi}, the total helicity flux is now written as the integral over the total 
magnetic flux crossing $\surf $ of the helicity flux density in each elementary flux 
tube {\it ``c''} that compose $\vol $. 
Then, by separating the contributions of helicity flux at $\xx_{c_{+}}$ from those 
at $\xx_{c_{-}}$, \inlinecite{Pariat05} expressed the helicity flux density per unit 
of magnetic flux tube, $\rmd h_{\Phi}/\rmd t$, as a field--weighted average of the 
flux per unit surface, $\gth$, at both footpoints $\xx_{c_{+}}$ and $\xx_{c_{-}}$ 
of flux tube {\it ``c''}: 
      \BE \label{eq:Eq-dh-gth}
         \deriv{h_{\Phi}}{t}\bigg|_{c} =  \frac{\gth(\xx_{c_{+}})}{|\Bn(\xx_{c_{+}})|} 
                                                             + \frac{\gth(\xx_{c_{-}})}{|\Bn(\xx_{c_{-}})|} \,.
      \EE

  %{\S}{\bf --- Gphi definition} \\

From \eq{Eq-dh-gth}, a helicity flux density per unit surface can be defined by 
redistributing the total helicity injected into flux tube {\it ``c''} at both footpoints of 
the flux tube with the fractions $f(\xx_{c_{+}})=f_{+}=f$ and $f(\xx_{c_{-}})=f_{-}=1-f$. 
They thus defined the best surface helicity flux density proxy, $\gph$, 
by equally sharing $\rmd h_{\Phi}/ \rmd t$ between the two footpoints of flux tube 
{\it ``c''}:
      \BE \label{eq:Eq-Gphi}
	\gph(\xx_{c_{\pm}}) = f_{\pm} \left( \gth(\xx_{c_{\pm}}) + 
	         \left| \frac{\Bn(\xx_{c_{\pm}})}{\Bn(\xx_{c_{\mp}})}\right| \gth(\xx_{c_{\mp}}) \right) \,,
      \EE
with $f_{+}=f_{-}=1/2$.
	
  %{\S}{\bf --- Using Gphi rather than Gtheta} \\

There is therefore a conceptual difference between $\gth$ and $\gph$. The 
proxy $\gph$ assumes that the footpoints of the elementary flux tubes have 
a knowledge of the helicity injection at the other footpoint. 
Therefore, when using $\gph$ one assumes that the helicity injection process is done with a 
characteristic timescale which is much larger than the transit time of 
information within the field line. As such information will be transferred 
through Alv\'enic waves along the field line, $\gph$ is meaningfull for any 
process which velocity is smaller than the averaged Alfv\'en speed along 
the field line. In the solar atmosphere such condition is easily satisfied as 
the typical velocities in the photosphere ($< 1\ \textrm{km \ s}^{-1}$) are orders 
of magnitude smaller than the coronal Alfv\'en speed ($\approx 10^3-10^4\ \textrm{km \ s}^{-1}$). 
The different motions that enable the energy storage in the coronal field 
are consistent with the use of $\gph$. However when considering 
processes occurring over the Alv\'enic timescale, 
such as magnetic reconnection, this condition may not be fulfilled.

\section{Methodology} \label{sec:S-methodo}

  %{\S}{\bf --- Goal} \\
  	
For the first time, we implement a general method to compute the helicity flux density, 
$\gph$. Our aim is to validate the method and study the properties of $\gph$ on case 
studies before applying it to observational studies. In order 
to interpret the $\gph$ maps we will obtain in observational studies, we need to 
know the helicity distribution associated to typical flux transport velocity fields and 
how the properties of the magnetic connectivity change these helicity distributions. 
We also need to know how the different parameters 
used for the $\gph$ computation can influence the results (\eg resolution of the 
magnetograms and $\gth$ maps, precision used for field lines integration).

  %{\S}{\bf --- Computing Gtheta and Gphi} \\

To compute $\gth$ from \eq{Eq-Gtheta}, we need the normal component of the 
magnetic field and the relative rotation rate of elementary magnetic 
flux tubes on $\surf $. The extra information needed to compute 
$\gph$ is the magnetic field in $\vol $.

\subsection{Flux Transport Velocity} \label{sec:S-Meth-FTV}

  %{\S}{\bf --- Chosen FTV} \\

\begin{figure} [t]   
   \centerline{\includegraphics[width=0.99\textwidth,clip=]{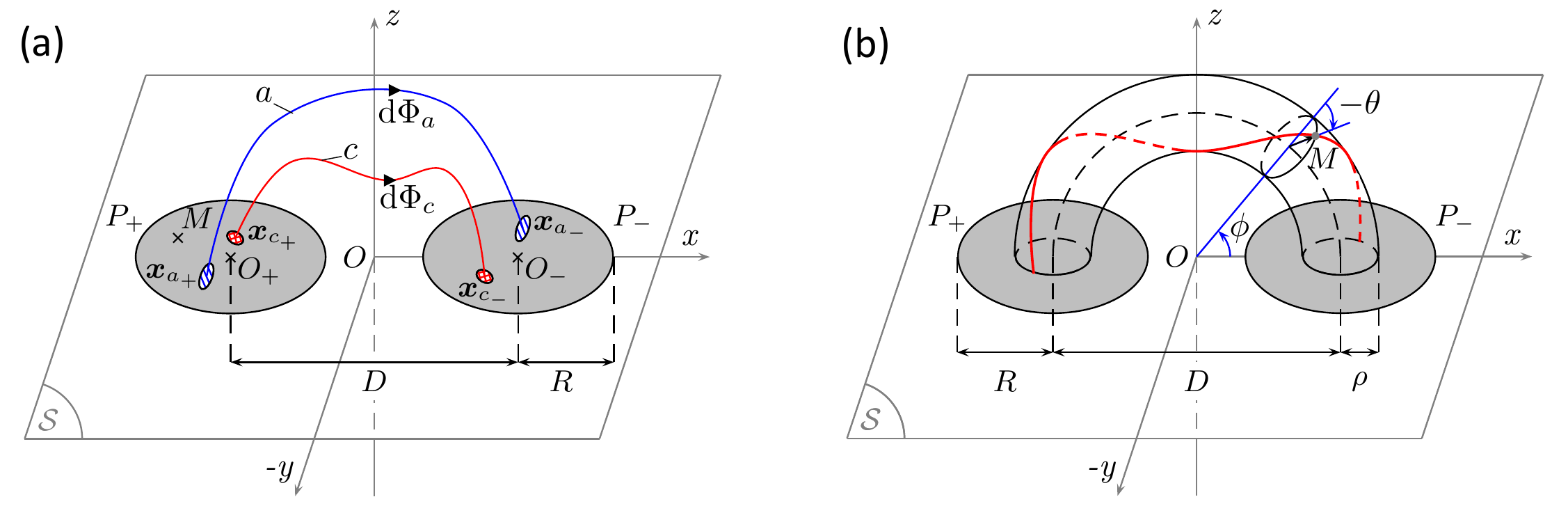}
              }
              \caption{(a) Sketch of the general magnetic configuration considered in our models constructed with two opposite and magnetic--flux balanced polarities, $P_{+}$ and $P_{-}$. The $\surf $--surface ($z=0$--plane) symbolizes the photosphere. The two magnetic field lines {\it ``a''} (blue line) and {\it ``c''} (red line) show the general connectivity. The magnetic field lines {\it ``a''} and {\it ``c''} go from magnetic footpoints $x_{a_{+}}$ and $x_{c_{+}}$ on $S$ to magnetic footpoints $x_{a_{-}}$ and $x_{c_{-}}$ with magnetic fluxes d$\Phi_{a}$ and d$\Phi_{c}$, respectively.. (b) Scheme of the torus configuration. The two polarities ($P_{+}$ and $P_{-}$) are the bases of the torus. The red line represents a magnetic field line of the torus at a radius $\rho$ from its axis.
                      }
   \label{fig:Fig-Methodo-connecti}
   \end{figure}

Observations (\eg \opencite{Moon02}; \opencite{Nindos03}; 
\opencite{Chae04}; \opencite{Schuck06}; \opencite{Jeong07}; 
\opencite{Labonte07}; \opencite{Welsch09}; \opencite{Jing12}) 
have reported complex patterns of photospheric 
flux transport velocities during the lifetimes of ARs involving the 
combination of separations and rotations of the entire or of parts of 
the magnetic polarities. 
In the following, we consider some elementary photospheric 
relative motions of two connected opposite magnetic polarities: two 
separating polarities without any rotation, a polarity rotating around 
another one, and two counter--rotating polarities. The polarities are 
isolated and magnetic--flux balanced. We consider a cartesian domain 
centered on point $O$ (see \fig{Fig-Methodo-connecti}). 
The positive ($P_{+}$) and negative ($P_{-}$) polarities are centered on 
$O_{+}$ and $O_{-}$ on the $x$-axis, respectively. The polarities are separated by 
the distance $D=| \vec{O}_{+} \vec{O}_{-} |$.

%:          Figure: Connectivity and torus schemes
%%%%%%%%%%%%%%%%%%%%%%%%%%%%%%%%%%%%%%%%%
%%%%%%%%%%%%%%%%%% FIGURE 1 %%%%%%%%%%%%%%%%%
%%%%%%%%%%%%%%%%%%%%%%%%%%%%%%%%%%%%%%%%%
%%%%%%%%%%%%%%%%%%%%%%%%%%%%%%%%%%%%%%%%%

  %{\S}{\bf --- FTV & Gtheta for the two separating polarities} \\

The considered flux transport velocity field at point $M(\xx)$ for the two separating polarities model 
is: 
     \BA  \label{eq:Eq-FTV-sepx}
     	\uu (M) &=&  \mp U_{0} {\vec e}_{x} \,, \quad \textrm{for} \ \pm \Bn (M)>0 \,,
      \EA
where $U_{0}$ is a positive constant. The associated helicity flux density $\gth$ is 
therefore (see \app{App-sepx} for a detailed derivation):
     \BE  \label{eq:Eq-Gth-sepx}
	  \gth (M) \ = \ \frac{U_{0} \Bn (M)}{\pi} \ \frac{\vec{O}_{\mp}\vec{M} \cdot \vec{e}_{y}}{|\vec{O}_{\mp}\vec{M}|^{2}} \ \Phi_{0} \,, \quad \textrm{for} \ \pm \Bn (M)>0 \,,
      \EE
where $\Phi_{0}$ is the absolute value of the total magnetic flux of each polarity.

  %{\S}{\bf --- FTV & Gtheta for the neg. pol. rigidly rotating around the pos.} \\

For the model of the negative polarity rigidly rotating around the positive one, the flux 
transport velocity field is:	
    \BA  \label{eq:Eq-FTV-onerot}
     	\uu (M) &=&  \vec{\Omega} \times {\vec O}_{+} \vec{M} \,, \quad \textrm{for} \ \Bn (M)<0 \,,
      \EA
where $\vec{\Omega}=\Omega \vec{e}_{z}$, $\Omega$ being the positive constant 
rotation rate of the negative polarity. From \app{App-onerot}, the associated helicity flux density 
is non--zero only if $M$ is in the positive polarity, and its expression is given by:
    \BE  \label{eq:Eq-Gth-onerot}
	\gth (M) \ = \ - \ \frac{\Omega \Bn (M)}{2\pi} \ \frac{\vec{O}_{-} \vec{M} \cdot \vec{O}_{+} \vec{O}_{-}}{|\vec{O}_{-} \vec{M}|^{2}} \ \Phi_{0} \,, \quad \textrm{for} \ \Bn (M)>0 \,.
      \EE

  %{\S}{\bf --- FTV & Gtheta for the two counter--rotating polarities} \\

For the third motion model --- \ie two counter--rotating opposite magnetic 
polarities --- the flux transport velocity field is:
    \BA  \label{eq:Eq-FTV-tworot}
     	\uu (M) &=&  \mp \vec{\Omega} \times {\vec O}_{\pm} \vec{M} \,, \quad \textrm{for} \ \pm \Bn (M)>0 \,,
      \EA
resulting in a helicity flux density (see \app{App-tworot}):
    \BE  \label{eq:Eq-Gth-tworot}
	\gth (M) \ = \ \mp \frac{\Omega \Bn (M)}{2\pi}\ \frac{\vec{O}_{\mp}\vec{M} \cdot \vec{O}_{+}\vec{O}_{-}}{|\vec{O}_{\mp}\vec{M}|^{2}}\ \Phi_{0} \,, \quad \textrm{for} \ \pm \Bn (M)>0 \,.
      \EE

\subsection{Magnetic Field} \label{sec:S-Meth-Bfield}

  %{\S}{\bf --- Chosen magnetic field} \\
  
Because we want to estimate the precision of the method, it is 
worthwhile to first consider simple analytical magnetic fields for which 
the connectivity is theoretically known. This allows us to estimate 
the precision of our computing method of $\gph$.

  %{\S}{\bf --- Two magnetic charges} \\

\begin{figure}
   \centerline{\includegraphics[width=0.99\textwidth,clip=]{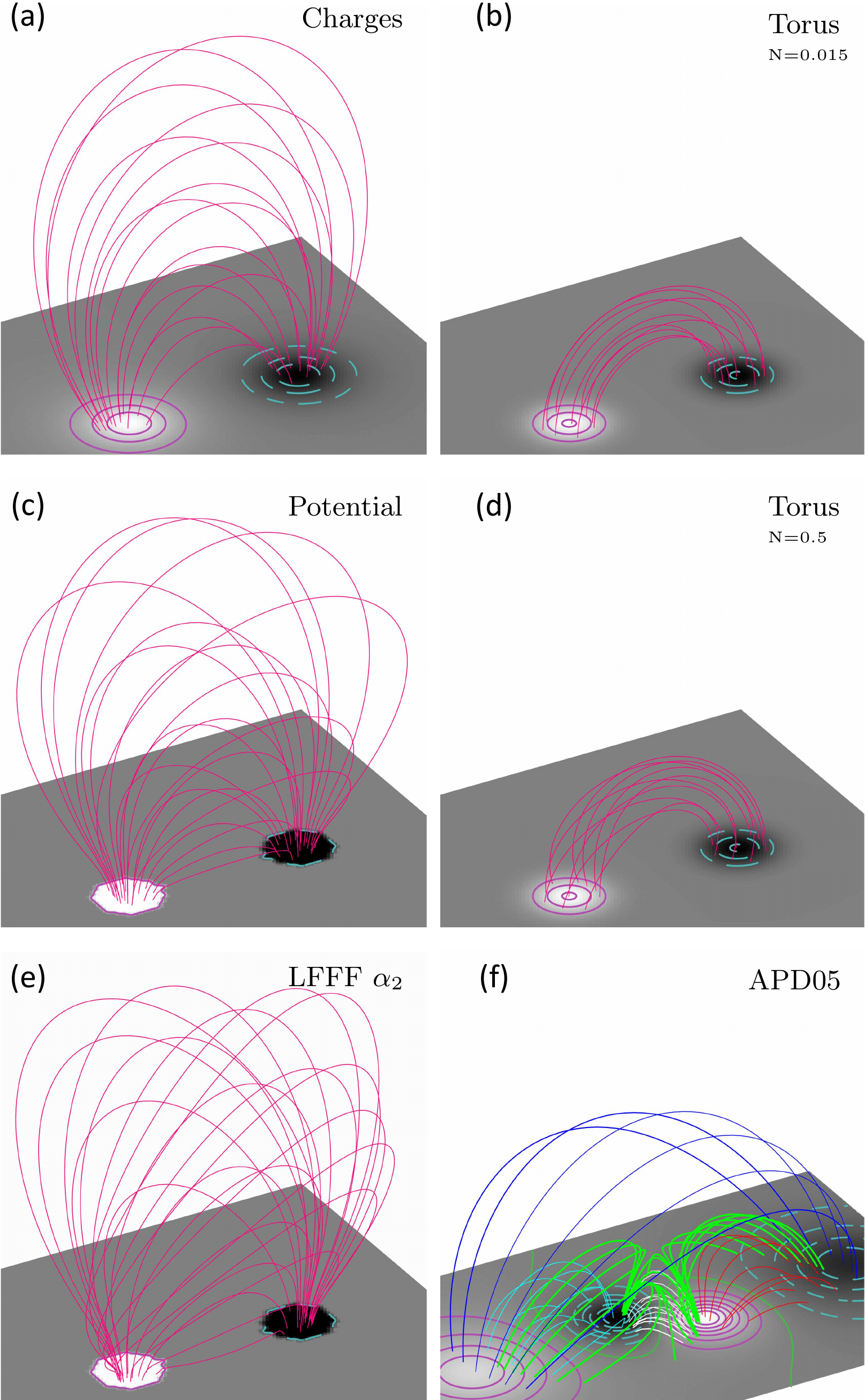}
              }
              \caption{Magnetic field configurations considered in our investigations. The magnetograms are represented at the $z=0$ plane with superposed isocontours of the magnetic field (cyan dashed and purple solid--lines for negative and positive values of the magnetic field respectively). For all but panel (f), the magnetic field lines are represented by the pink lines and the magnetic field values are between $1000 \ \textrm{Gauss}$ (white) and $-1000 \ \textrm{Gauss}$ (black). 
   (a) Two magnetic charges. 
   (b,d) Half--emerged torus with a twist $N=0.015$ and $N=0.5$ respectively. 
   (c,e) Two uniform opposite magnetic polarities with linear constant--$\alpha$ force--free field extrapolations $\alpha=0$ and $\alpha=5.6 \times 10^{-3} \ \textrm{Mm}^{-1}$ respectively. 
   (f) MHD simulation of a magnetic configuration with quasi--separatrix layers (QSLs). Different colors (red, white, cyan and blue) have been used to show the different quasi--domains of connectivity. Thick green magnetic field lines correspond to QSLs field lines (see \sect{S-Hinj-in-QSLs}). The S--like green line on the magnetogram corresponds to the polarity inversion line.  
                      }
   \label{fig:Fig-Bfields}
   \end{figure}

One simple analytical field to start with is a potential 
magnetic field (see \fig{Fig-Bfields}a). Such a field is constructed by 
placing two artificial opposite magnetic charges below the photosphere. 
The positive and negative magnetic charges are placed at $A_{+}$ and $A_{-}$, 
resulting in:
    \BE  \label{eq:Eq-Bcharges}
	\bb (\xx)=q_{0} \frac{\xx-\xx_{+}}{|\xx-\xx_{+}|^{3}}-q_{0} \frac{\xx-\xx_{-}}{|\xx-\xx_{-}|^{3}} \,,
      \EE
where $\xx_{\pm} = \vec{O} \vec{A}_{\pm}$ and $q_{0}$ is the absolute value of the 
positive and negative magnetic charges.

  %{\S}{\bf --- Half--emerged torus} \\

Theoretical and numerical simulations studies 
have shown that, to emerge into the corona, a magnetic flux tube needs 
some twist \cite{Emonet98}. Observational studies 
also highlight that as an AR appears, important amounts of helicity are injected 
into the corona (\eg \opencite{Chae01}; \opencite{Kusano04b}; \opencite{Pariat06}; 
\opencite{Jeong07}; \opencite{Romano11a}; \opencite{Jing12}), with evidences of 
a twisted flux tube \cite{Luoni11}. Therefore, we consider a second magnetic 
field defined by a uniformly twisted torus half--emerged 
into the corona (see panels b and d of \fig{Fig-Bfields}; \opencite{Luoni11}). 
The associated magnetic field is 
$\bb=B_{\theta} \vec{e}_{\theta} + B_{\phi} \vec{e}_{\phi}$, such that:
    \BA
     	B_{\theta} (M) &=& \frac{4 N \rho}{D + 2 \rho cos(\theta)} \ B_{\phi} (M) \,,  \nonumber \\
	&& \nonumber \\
	B_{\phi} (M)    &=& -B_{0} e^{-\left(\rho / R \right)^{2}}  \,,
	\label{eq:Eq-Btorus}
      \EA
where $N$ corresponds to the number of turns of the magnetic field lines around 
the torus axis in half the torus, and $B_{0}$ is the magnetic field strength at 
the center of the positive polarity. The torus center is located at photospheric 
point $O$, and the distance $D/2$ defines the main radius of the torus. 
The variables $\rho$, $\theta$ and $\phi$ are 
respectively the distance to the torus axis, the rotation angle around the torus axis, 
and the location angle along the torus axis (see \fig{Fig-Methodo-connecti}b).

%:          Figure: Magnetic field configurations
%%%%%%%%%%%%%%%%%%%%%%%%%%%%%%%%%%%%%%%%%
%%%%%%%%%%%%%%%%%% FIGURE 2 %%%%%%%%%%%%%%%%%
%%%%%%%%%%%%%%%%%%%%%%%%%%%%%%%%%%%%%%%%%
%%%%%%%%%%%%%%%%%%%%%%%%%%%%%%%%%%%%%%%%%   

  %{\S}{\bf --- Linear force--free fields} \\
  
For the two previous cases, the magnetic field is analytical. However, 
extrapolations of observed magnetograms are in most cases non--analytical 
fields given on a discrete mesh. The consequence is that errors due to 
extrapolations and interpolations of the magnetic field will possibly degrade or modify 
the signal in $\gph$ maps. To investigate it, we performed linear force--free field extrapolations 
(see \figpnls{Fig-Bfields}{c}{e}) of a magnetogram defined by:
    \BA  \label{eq:Eq-Buniform}
     	\bb_{z=0} (M) &=& \pm B_{0} \vec{e}_{z} \,, \quad \textrm{for} \ M \ \textrm{in} \ P_{\pm} \,,
      \EA
where $B_{0}$ is the magnetic field strength in the positive polarity and, 
$P_{+}$ and $P_{-}$ refer to the positive and negative 
magnetic polarities which are circular of radius $R$ (cf \fig{Fig-Methodo-connecti}a). 
The extrapolations were performed using the code XTRAPOL 
(\eg \opencite{Amari99}; \opencite{Amari06}; \opencite{Amari10}). The code 
solves the Poisson's equation for the vector potential, $\grad^{2} \avec +\alpha^2 \avec=0$, with 
the boundary conditions given by \eq{Eq-Buniform} inside the photospheric 
polarities and $\bb \cdot \rmd \surf =0$ elsewhere on the 
boundaries of the extrapolation domain. The vector potential formulation 
ensures that the solenoidal condition is verified to the 
machine precision. The boundary conditions assumes that no field 
go through the lateral and top boundaries. This imposes that the total magnetic flux 
within the positive polarity is entirely connected to the negative polarity. 
We make this choice to prevent the presence 
of open magnetic field lines in the domain in order to compute $\gph$ 
for all photospheric magnetic footpoints of $\gth$ maps. 
Note also that, the spatial resolution of the extrapolated fields is chosen to 
be different from the one used to compute $\gth$ (see the fragmented 
shape of the magnetic field isocontours in \figpnls{Fig-Bfields}{c}{e}). 
We make this choice to investigate the effect of using a different 
resolution between the extrapolations and the helicity flux density 
maps. In practice, we find no significant effects as long as the difference of 
resolution is lower than $8$ for our high resolution helicity flux density maps. This 
means that we can use a lower resolution in the extrapolation without modifying 
the resulting $\gph$ map.

  %{\S}{\bf --- Quadrupolar configurations} \\

The last magnetic configuration considered is more complex as it involves 
quasi--separatrix layers in MHD simulations. This magnetic field and the 
associated flux transport velocity fields will be described in \sect{S-Hinj-in-QSLs}.

  %{\S}{\bf --- Note on Gtheta/Gphi maps delimited to two circular polarities} \\

%:          Figure: Polarity to subpolarity
%%%%%%%%%%%%%%%%%%%%%%%%%%%%%%%%%%%%%%%%%
%%%%%%%%%%%%%%%%%% FIGURE 3 %%%%%%%%%%%%%%%%%
  \begin{figure} [b]   
   \centerline{\includegraphics[width=0.79\textwidth,clip=]{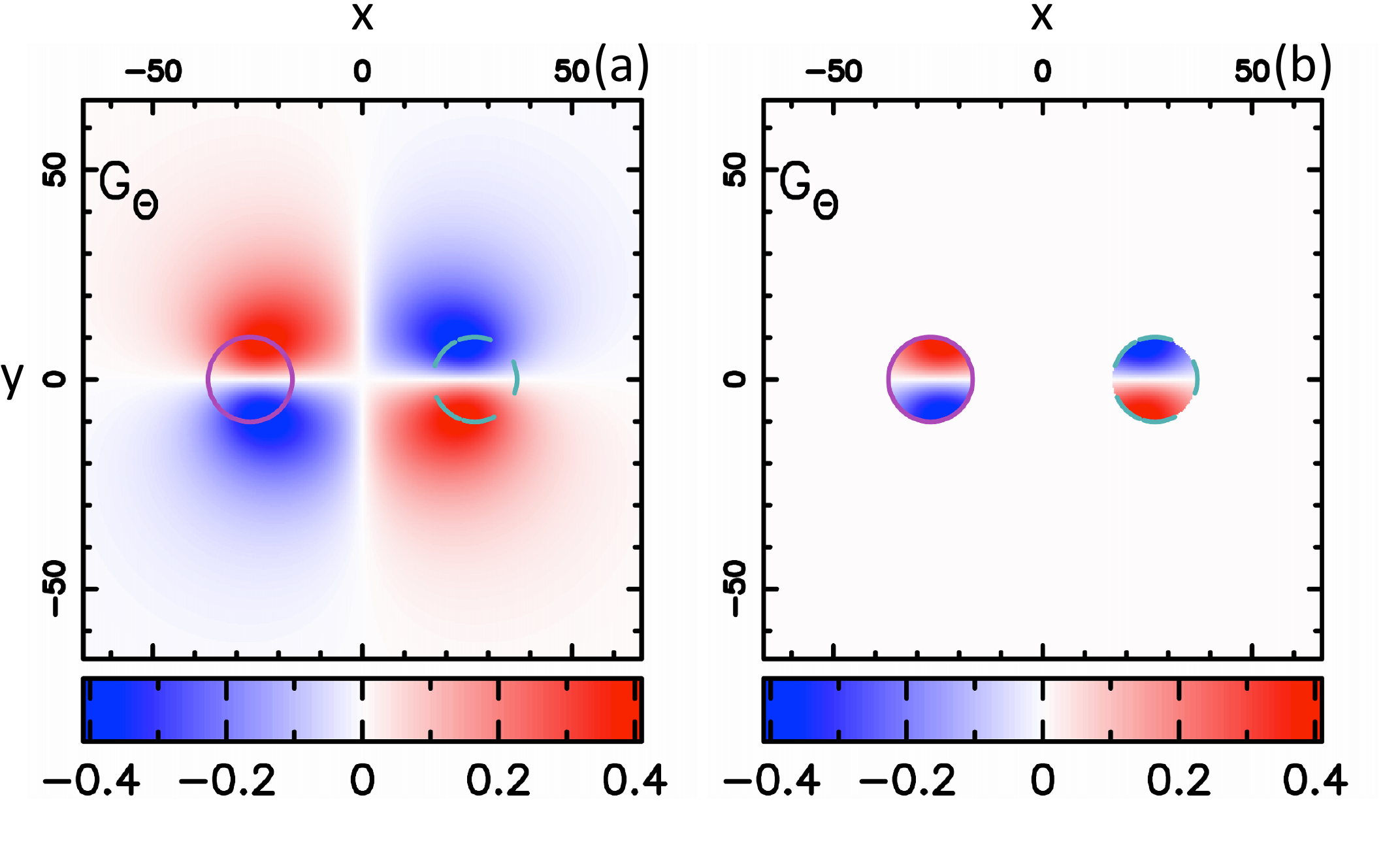}
              }
              \caption{
   (a) Full $\gth$ map of the analytical potential magnetic field (\eq{Eq-Bcharges}). 
   (b) $\gth$ map for the considered polarities with positive (solid purple) and negative (dashed cyan) isocontours of the magnetic field delimiting these two polarities.
                      }
   \label{fig:Fig-Poltosubpol}
   \end{figure} 
%%%%%%%%%%%%%%%%%%%%%%%%%%%%%%%%%%%%%%%%%
%%%%%%%%%%%%%%%%%%%%%%%%%%%%%%%%%%%%%%%%%

For the extrapolated magnetic fields, the magnetic flux is restricted 
to two circular regions (as defined in 
\fig{Fig-Methodo-connecti}a). However, this is not strictly the case for the two 
analytical magnetic fields: even if the field strength strongly decreases away 
from $O_{\pm}$, it does not vanish. Consequently, there is helicity 
signal in the whole domain considered in the $\gth$ and $\gph$ maps 
for the analytical magnetic fields of \eqs{Eq-Bcharges}{Eq-Btorus} 
as illustrated for $\gth$ in \fig{Fig-Poltosubpol}a. 
For coherence of the results, we always extract and consider the 
helicity flux signal from the two connected polarities of radius R as shown 
in \fig{Fig-Poltosubpol}b, and which contain most of the magnetic flux. 
From now on, the positive and negative polarities 
$P_{+}$ and $P_{-}$ will refer to these two polarities, and all scalings of 
\sect{S-Meth-setup} are made with respect to them. 
However, it must be emphasized that, in all our models, $\gph$ 
computation takes into account the motion of all the magnetic flux at $z=0$.

\subsection{Numerical Setup} \label{sec:S-Meth-setup}

  %{\S}{\bf --- Parameters values} \\

All maps are computed in a cartesian domain with $400 \times 400$ points. The scaling 
is chosen to take into account the typical values obtained from observations. The 
$z=0$--plane, which represents the photosphere, covers the $xy$-domain 
$[-67,67] \times [-67,67] \ \textrm{Mm}^{2}$. The centers of the photospheric polarities 
are separated by the distance $D=54 \ \textrm{Mm}$. The positive and the negative 
magnetic charges are placed at points $A_{+}$ of coordinates $(-27,0,-13)$ Mm 
and $A_{-}$ of coordinates $(27,0,-13)$ Mm, respectively. The radius of the 
photospheric polarities is set to $R=10 \ \textrm{Mm}$. The maximum value of the 
normal component of the magnetic field, $B_{0}$, and of the flux transport velocity 
fields, $U_{0}$, are set to $1000 \ \textrm{Gauss}$ and $0.1 \ \textrm{km \ s}^{-1}$ respectively. 
The extrapolations were performed on a non--uniform mesh covering the domain 
$[-533,533] \times [-533,533] \times [0,1066] \ \textrm{Mm}^{3}$ with 
$513 \times 513 \times 200$ points.

  %{\S}{\bf --- Scaling} \\

The above scalings lead to helicity flux densities in units of 
$10^{6} \ \textrm{Wb}^{2} \textrm{.m}^{-2} \textrm{.s}^{-1}$, and a total helicity flux 
in units of $10^{21} \ \textrm{Wb}^{2} \textrm{.s}^{-1}$ which are typical observed values 
in ARs (\eg \opencite{Chae01}; \opencite{Chandra10}).

  %{\S}{\bf --- Computing gtheta and gphi} \\

$\gth$ maps are computed using \eqss{Eq-FTV-sepx}{Eq-Buniform}. Let us 
first consider the positive magnetic polarity ($\Bn > 0$). In practice, each 
$\gth$--mesh point $\xx_{a_{+}}$, is identified as the cross--section of an elementary 
magnetic flux--tube with the photosphere and is associated to the surface helicity flux 
density $\gth (\xx_{a_{+}})$ at this point. To compute 
$\gph (\xx_{a_{+}})$, we need the position of $\xx_{a_{-}}$ --- the second 
footpoint of the elementary magnetic flux--tube {\it ``a''} --- and its 
associated surface helicity flux density. 
Each elementary magnetic 
flux--tube is thus associated to one magnetic field line that is integrated to get 
the connectivity.
The integration is performed starting from 
$\xx_{a_{+}}$ to $\xx_{a_{-}}$ using the Fortran NAG--routine D02CJF, with the 
precision of the integration defined as $10^{-n}$. 
Thus, the higher $n$ is, the more precisely the 
connectivity and $\gph$ are computed. Generally, the $\xx_{a_{-}}$ footpoint 
does not fall on a mesh point. Thus, the values of $\gth$ and $\Bn$ at this 
point are bilinearly interpolated using the values at the four closest 
surrounding mesh points. If $\xx_{a_{-}}$ is not found on the ($z=0$)--plane 
(\eg open magnetic field lines), the value of $\gph$ at $\xx_{a_{+}}$ is 
simply set to $\gth (\xx_{a_{+}})$. Finally, the same procedure as above 
is used in the negative magnetic polarity ($\Bn < 0$) starting the magnetic field 
line integration from $\xx_{a_{-}}$.

%To compute $\gph$ and the connectivity of footpoint $\xx_{a_{+}}$ on $\gth$--mesh, the magnetic field line {\it ``a''} is integrated starting from $\xx_{a_{+}}$ to $\xx_{a_{-}}$ using the Fortran NAG--routine D02CJF. The precision of the integration is defined as $10^{-n}$. Thus, the higher $n$ is, the more precisely the connectivity and $\gph$ are computed. Generally, the $\xx_{a_{-}}$ footpoint does not fall on a mesh point. Thus, the values of $\gth$ and $\Bn$ at this point are bilinearly interpolated using the values at the four closest surrounding mesh points. If $\xx_{a_{-}}$ is not found on the ($z=0$)--plane (\eg open magnetic field lines), the value of $\gph$ at $\xx_{a_{+}}$ is simply set to $\gth (\xx_{a_{+}})$. Finally, the same procedure as above is used starting the integration from $\xx_{a_{-}}$.

\section{Results for Two Magnetic Charges} \label{sec:S-two-source-charges}
	
%\subsection{Two magnetic \e{source} charges} \label{sec:S-two-source-charges}
	
In this section, the magnetic field is given by \eq{Eq-Bcharges} for all 
flux transport velocity models and the associated magnetogram at the 
$z=0$--plane is displayed in \fig{Fig-Bfields}a.
	
\subsection{Two Separating Magnetic Polarities} \label{sec:S-Charges-sepx}	

\begin{figure}[h]
   \centerline{\includegraphics[width=0.99\textwidth,clip=]{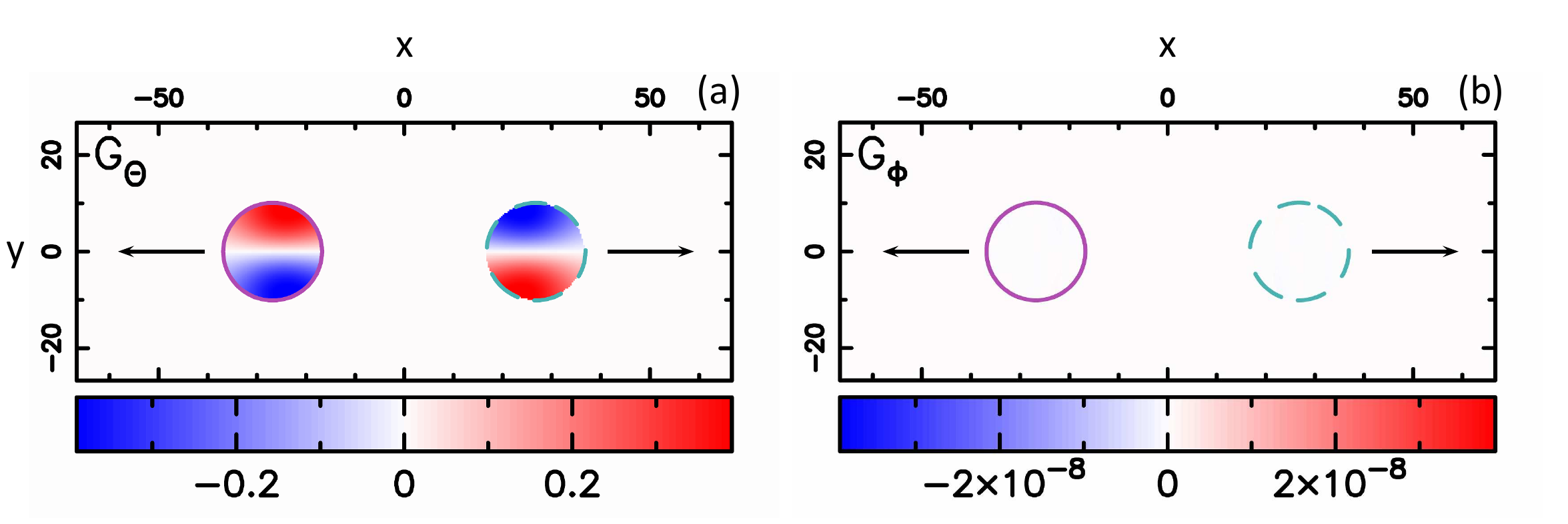}
              }
              \caption{Helicity flux density distribution, in $10^{6}\ \textrm{Wb}^{2} \textrm{m}^{-2} \textrm{s}^{-1}$, for the case of two separating opposite magnetic polarities without rotation. (a) $\gth$ map. (b) $\gph$ map. Solid purple and dashed cyan are isocontours of the magnetic field. Black arrows show the motion applied to the polarities. The saturation levels of $\gth$ and $\gph$ are different by eight orders of magnitude.
                      }
   \label{fig:Fig-Charges-sepx}
   \end{figure}

In this example, the two connected opposite magnetic polarities 
separate away from each other in the $x$-direction (see \eq{Eq-FTV-sepx}).
Since the polarities simply separate without any rotation, no 
helicity is injected to the system. 
However, as the two polarities separate, every elementary polarity sees 
a relative rotation of all other elementary polarities of opposite sign. 
This induces net non--zero values of $\gth$ as shown in the left panel of 
\fig{Fig-Charges-sepx}a. 

In this model, the symmetry of the magnetic field and of the applied velocity 
field implies $\gth(\xx_{a_{-}})=-\gth(\xx_{a{+}})$. 
Therefore, by taking the connectivity into account, $\gph$ is null to the numerical 
errors over all the two polarities (\fig{Fig-Charges-sepx}b). 
This simple example reveals the limits of $\gth$ to present a truthful 
distribution of helicity flux while $\gph$ gives the expected results.

%:          Figure: Helicity flux density - Charges separation
%%%%%%%%%%%%%%%%%%%%%%%%%%%%%%%%%%%%%%%%%
%%%%%%%%%%%%%%%%%% FIGURE 4 %%%%%%%%%%%%%%%%%
%%%%%%%%%%%%%%%%%%%%%%%%%%%%%%%%%%%%%%%%%
%%%%%%%%%%%%%%%%%%%%%%%%%%%%%%%%%%%%%%%%%   

\subsection{One Polarity Rigidly Rotating Around the Other} \label{sec:S-Charges-onerot}

%:          Figure: Helicity flux density - Charges rigid rotation
%%%%%%%%%%%%%%%%%%%%%%%%%%%%%%%%%%%%%%%%%
%%%%%%%%%%%%%%%%%% FIGURE 5 %%%%%%%%%%%%%%%%%
  \begin{figure}[b]
   \centerline{\includegraphics[width=0.99\textwidth,clip=]{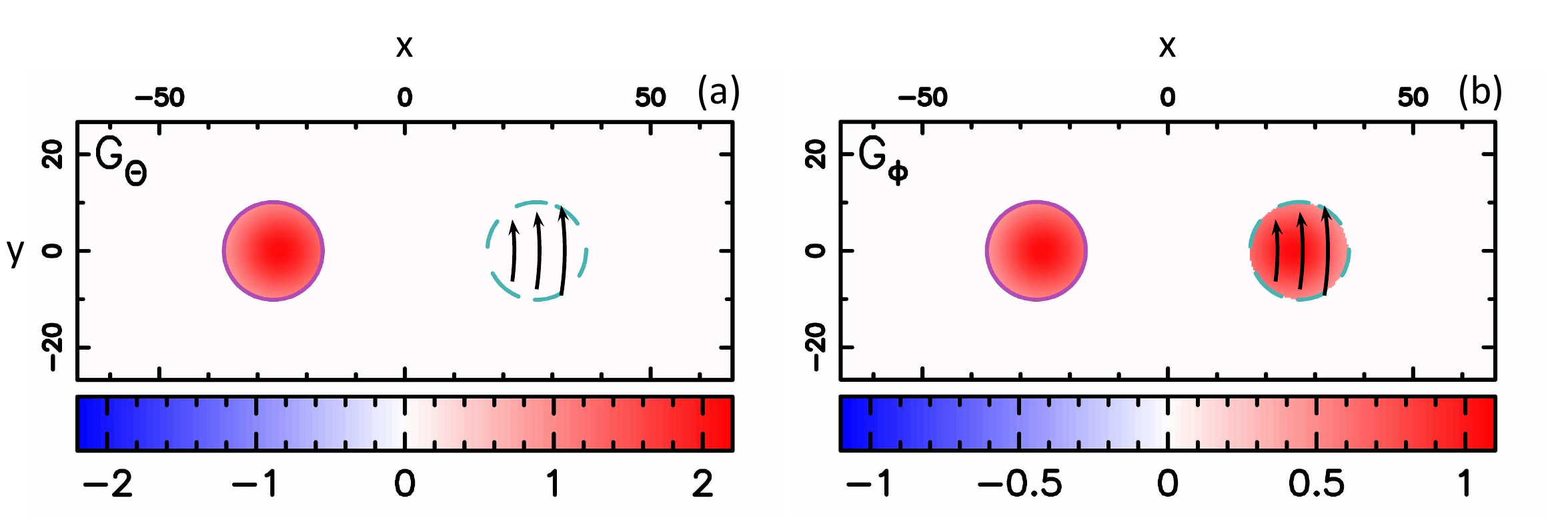}
              }
              \caption{Helicity flux density distribution, in $10^{6}\ \textrm{Wb}^{2} \textrm{m}^{-2} \textrm{s}^{-1}$ for the negative polarity rigidly rotating around the positive polarity. The drawing convention is the same as in \fig{Fig-Charges-sepx} (with a different color scale).
                      }
   \label{fig:Fig-Charges-onerot}
   \end{figure} 
%%%%%%%%%%%%%%%%%%%%%%%%%%%%%%%%%%%%%%%%%
%%%%%%%%%%%%%%%%%%%%%%%%%%%%%%%%%%%%%%%%%   

The positive polarity is fixed while the negative polarity rigidly rotates around 
$O_{+}$ (the center of the positive polarity, see \eq{Eq-FTV-onerot}). 
From \eq{Eq-Gth-onerot}, $\gth$ is non--zero only in the 
positive polarity. The reason is that, in \eq{Eq-Gtheta}, 
there are two terms that contribute to $\gth$ inside one polarity. 
One term is the relative motion of footpoints inside the polarity. The second is 
the relative motion with regard to the footpoints of the other polarity. In the negative polarity, the 
two terms cancel out (\app{App-onerot}). In the positive polarity, however, only the term coming 
from the relative motion of the negative polarity is non--zero.
	
This second example also presents the limits of $\gth$ maps to well 
localize the injection of helicity. This can be misleading when 
relating the injection of helicity to magnetic activity (\eg \opencite{Chandra10}). 
This is corrected in the corresponding $\gph$ map that shows that positive 
helicity is redistributed in both polarities (\fig{Fig-Charges-onerot}b).

\subsection{Two Counter--rotating Polarities} \label{sec:S-Charges-tworot}
	
Let us now consider the model of two counter--rotating polarities (\eq{Eq-FTV-tworot}). 
The positive and negative polarities rigidly rotate clockwisely and counterclockwisely
around their centers $O_{+}$ and $O_{-}$, respectively. 
This configuration illustrates the difference of 
assumptions in the definition of $\gth$ and $\gph$.

Indeed, if the rotation is slow enough, the system is equivalent to a 
non-twisted flux tube rotating around its central axis. With such driving, 
a non-twisted flux tube would appear similarly untwisted at any time. 
Thus, overall no helicity is injected to the system. One therefore expects that 
$\gph$ would correspond to this null injection of magnetic helicity.

\begin{figure}[b]
   \centerline{\includegraphics[width=0.99\textwidth,clip=]{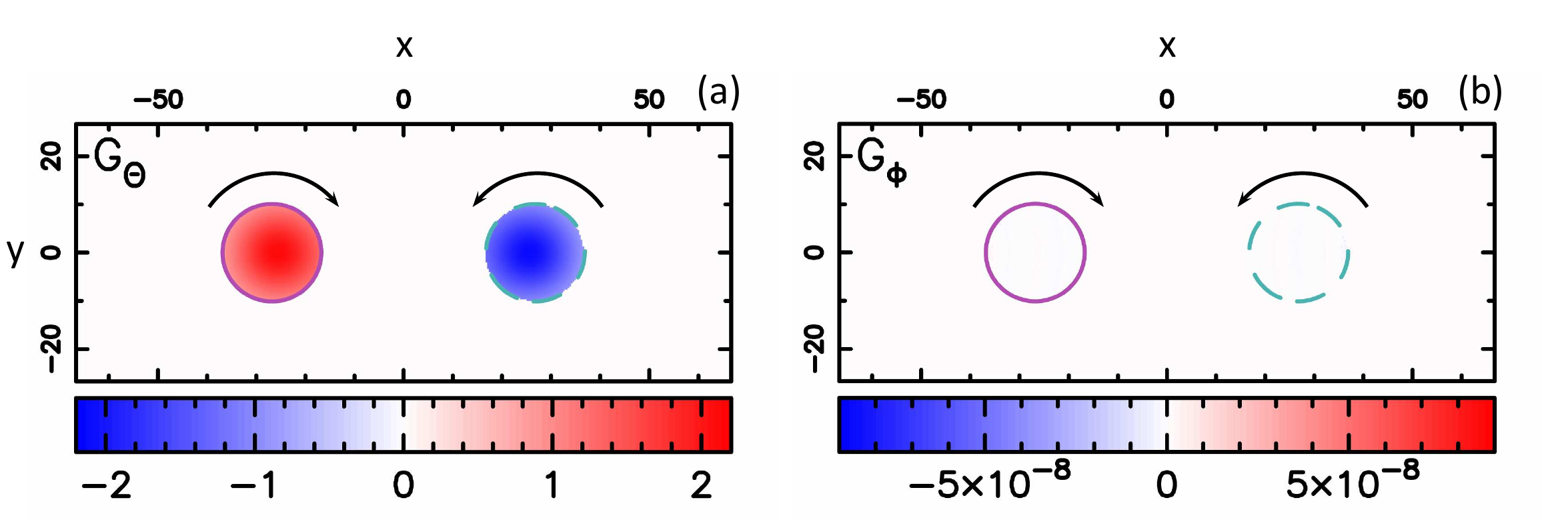}
              }
              \caption{Helicity flux density distribution, in $10^{6}\ \textrm{Wb}^{2} \textrm{m}^{-2} \textrm{s}^{-1}$ for the two counter--rotating opposite magnetic polarities. The drawing convention is the same as in \fig{Fig-Charges-sepx} (with a different color scale).
                      }
   \label{fig:Fig-Charges-tworot}
   \end{figure}

The $\gth$ map presents a distribution of helicity which is far different from 
a null injection (\fig{Fig-Charges-tworot}a). Taking the connectivity into 
account, \ie using $\gph$ (\fig{Fig-Charges-tworot}b), 
removes this helicity flux signal (regardless of numerical errors) allowing 
to get the expected null distribution of helicity flux.

%:          Figure: Helicity flux density - Charges counter--rotation
%%%%%%%%%%%%%%%%%%%%%%%%%%%%%%%%%%%%%%%%%
%%%%%%%%%%%%%%%%%% FIGURE 6 %%%%%%%%%%%%%%%%%
%%%%%%%%%%%%%%%%%%%%%%%%%%%%%%%%%%%%%%%%%
%%%%%%%%%%%%%%%%%%%%%%%%%%%%%%%%%%%%%%%%%   

While $\gth$ clearly misrepresents the global slow injection of helicity in 
this case, it would properly represent the helicity injection if the considered 
motion was extremely fast. Indeed if one considers counter rotating motions 
at a speed higher than the Alfv\'enic transit time, the opposite footpoint would 
have no indication of the helicity injection at the other footpoint. At the beginning 
of the injection, an initially untwisted flux rope would be such that 
the central part would stay untwisted but with oppositely twisted field around 
each footpoint. These counter--rotating motions would 
correspond to the launch of two rotating Alfv\'en waves of opposite sign. 
The $\gth$ maps do properly represent such helicity injection.  As time goes, 
these Alfv\'en waves would eventually cancel each other resulting in a null 
helicity budget for the system. For longer timescales, the $\gph$ map therefore 
better represents the proper helicity injection in the system.

\subsection{Errors Estimation} \label{sec:S-Charges-error}
	
In this section, we investigate the role of the parameter {\it n} 
--- used for field lines integration --- in the above $\gph$ maps. $\gth$ and the 
connectivity are analytically known which allows us to compute the 
theoretical value, $G_{\Phi,\rm{th}}$. Then, we estimate the error between the 
computed $\gph$ map from our numerical method and $G_{\Phi,\rm{th}}$ by 
computing the root mean square of $\gph-G_{\Phi,\rm{th}}$.

With the analytical magnetic field considered in this section, 
the resolution on the magnetic field is only limited 
by the computing precision, \ie $10^{-16}$ as the magnetic field was 
computed with a double precision. The numerical precision on $\gph$ 
is thus limited by the precision of:
	\begin{description}
		\item \hspace{1cm} $\cdot$ $\gth$, ${\rm{err}}_{\gth} \approx 10^{-16}$,
		\item \hspace{1cm} $\cdot$ the field lines integration, ${\rm{err}}_{{\rm{fli}}} \approx 10^{-n}$,
		\item \hspace{1cm} $\cdot$ the computation of $\Bn$ at the second footpoint, ${\rm{err}}_{\Bn} \approx 10^{-16}$,
		\item \hspace{1cm} $\cdot$ the bilinear interpolation of $\gth$ at the second footpoint, ${\rm{err}}_{{\rm{interp}}}$.
	\end{description}

The total error at each mesh point, ${\rm{err}}_{{\rm{tot}}}$, can thus be estimated as follows: 
       \BE \label{eq:Eq-guesstimated-error}
       	   {\rm{err}}_{{\rm{tot}}} \approx \sqrt{{\rm{err}}_{\gth}^{2}+{\rm{err}}_{{\rm{fli}}}^{2}+{\rm{err}}_{\Bn}^{2}+{\rm{err}}_{{\rm{interp}}}^{2}}   \,.
       \EE
For smooth variations of $\gth$ (as in all our cases), the error from the 
bilinear interpolation should not be the most limiting one. In this case, 
for $n<16$, the precision on $\gph$ is expected to be limited by the precision of 
field line integration. The consequence is that we expect an exponential 
decrease of the error as $n$ increases.

%:          Figure: Charges - Gphi errors estimation
%%%%%%%%%%%%%%%%%%%%%%%%%%%%%%%%%%%%%%%%%
%%%%%%%%%%%%%%%%%% FIGURE 7 %%%%%%%%%%%%%%%%%
  \begin{figure}[t]
   \centerline{\includegraphics[width=0.59\textwidth,clip=]{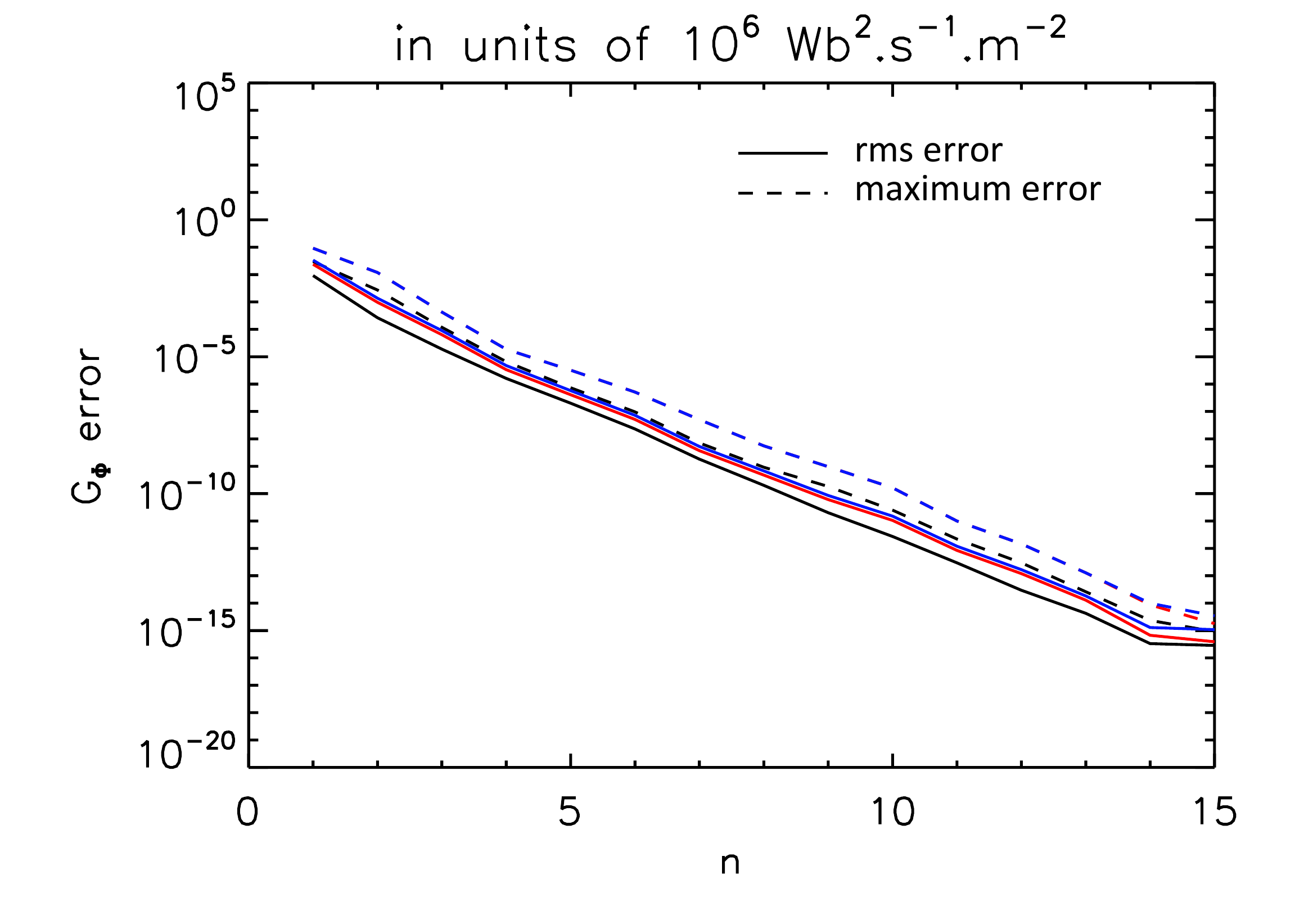}
              }
              \caption{Computed root mean square (solid lines) and maximum error (dashed lines) from $\gph$ maps for two separating polarities (black), the negative polarity rigidly rotating around the positive one (red), and two counter--rotating magnetic polarities (blue), as a function of the field line integration parameter $n$ defining the numerical integration precision ($=10^{-n}$).
                      }
   \label{fig:Fig-Charges-error}
   \end{figure} 
%%%%%%%%%%%%%%%%%%%%%%%%%%%%%%%%%%%%%%%%%
%%%%%%%%%%%%%%%%%%%%%%%%%%%%%%%%%%%%%%%%%   

\fig{Fig-Charges-error} displays the influence of {\it n} on 
the root mean square and the maximum error of $\gph$ maps. 
As expected, the figure shows that both $\gph$ 
rms and maximum error exponentially decrease as $n$ increases. 
Therefore, the precision on $\gph$ is indeed limited by the 
precision of field lines integration. The rms on 
$\gph$, for the three cases shown \figss{Fig-Charges-sepx}{Fig-Charges-tworot} 
($n=8$), is $2$, $5$ and $7 \times 10^{-10}$ $\gth$ units respectively, 
\ie more than $10^{9}$ times smaller than the typical values of the signal 
found in $\gth$ maps. Hence, our numerical method allows us to compute the 
distribution of helicity flux, $\gph$, with a very good accuracy.

\section{Results for a Half Emerged Torus} \label{sec:S-Torus}
	
In this section, the magnetic field is a uniformly twisted torus half 
emerged into the solar corona  (see \figpnls{Fig-Bfields}{b}{d} and \eq{Eq-Btorus}; 
\opencite{Luoni11}). The two opposite magnetic polarities are thus the 
two intersections of the torus with the photosphere (\fig{Fig-Methodo-connecti}b).

The amount of helicity, $H$, found in ARs can be converted to a uniform 
twist, $N^\prime$, with $H=N^\prime \Phi^{2}$, where $\Phi$ is the AR 
magnetic flux (average of both polarities). Observations report typical 
values of $N'$ from $\approx 0.01$ to $\approx 0.3$ (\opencite{Demoulin09} and 
references therein). In the following, we thus consider the torus configuration 
with $N=0.015$, $0.05$, and $0.5$.

Note that, the case $N=0$ has the same type of connectivity as the two 
magnetic charges but with a different $\Bn$ distribution. 
Therefore, we expect similar helicity flux distributions as in 
\figss{Fig-Charges-sepx}{Fig-Charges-tworot} when the same velocity models 
are applied.

\subsection{Two Separating Magnetic Polarities}	 \label{sec:S-Torus-sepx}

\begin{figure}[h]
   \centerline{\includegraphics[width=0.99\textwidth,clip=]{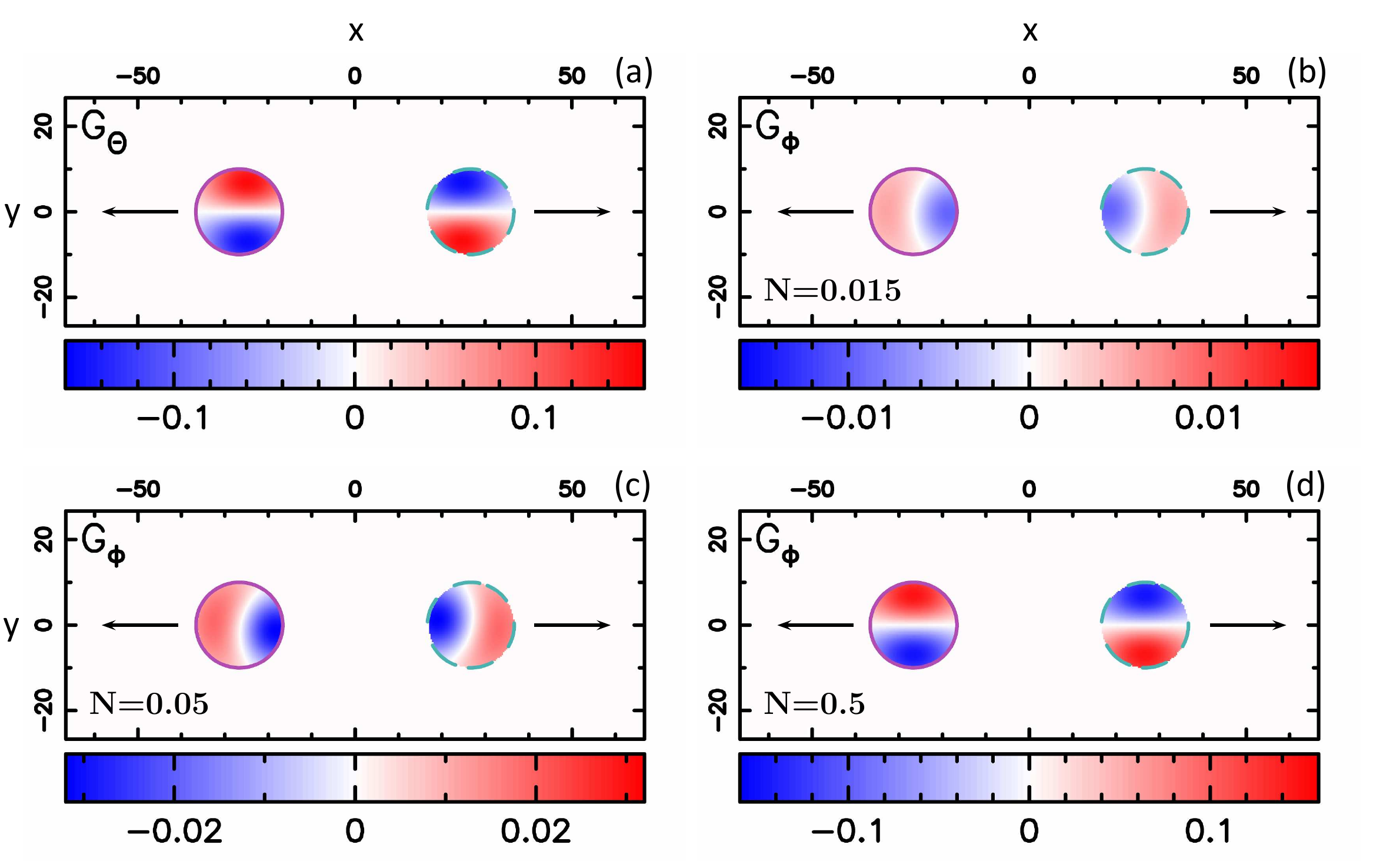}
              }
              \caption{$\gth$ and $\gph$ maps for the torus magnetic field configuration with two separating magnetic polarities. (a) $\gth$ map. (b,c,d) $\gph$ maps for a twist $N=0.015$, $0.05$ and $0.5$ respectively. The drawing convention is the same as in \fig{Fig-Charges-sepx} (notice the different color scales).
                      }
   \label{fig:Fig-Torus-sepx}
   \end{figure}

As for the potential magnetic field of \sect{S-Charges-sepx}, the two 
polarities separate without any rotation (flow given by 
\eq{Eq-FTV-sepx}) implying that no helicity is 
injected to the system. As expected, $\gth$ (\fig{Fig-Torus-sepx}a) exhibits 
a similar distribution as for the case with two magnetic charges (\fig{Fig-Charges-sepx}a), 
and the total helicity flux computed from $\gth$ and $\gph$ maps is indeed zero 
(\sect{S-Charges-sepx}). 
However, as shown by \fig{Fig-Torus-sepx}b-d, $\gph$ maps also present helicity injection with 
both signs of helicity in both polarities. But are these signals 
in $\gph$ maps real, or are they spurious signals as in the $\gth$ map?

%:          Figure: Helicity flux density - Torus separation
%%%%%%%%%%%%%%%%%%%%%%%%%%%%%%%%%%%%%%%%%
%%%%%%%%%%%%%%%%%% FIGURE 8 %%%%%%%%%%%%%%%%%
%%%%%%%%%%%%%%%%%%%%%%%%%%%%%%%%%%%%%%%%%
%%%%%%%%%%%%%%%%%%%%%%%%%%%%%%%%%%%%%%%%%   

\begin{figure}[h]
   \centerline{\includegraphics[width=0.99\textwidth,clip=]{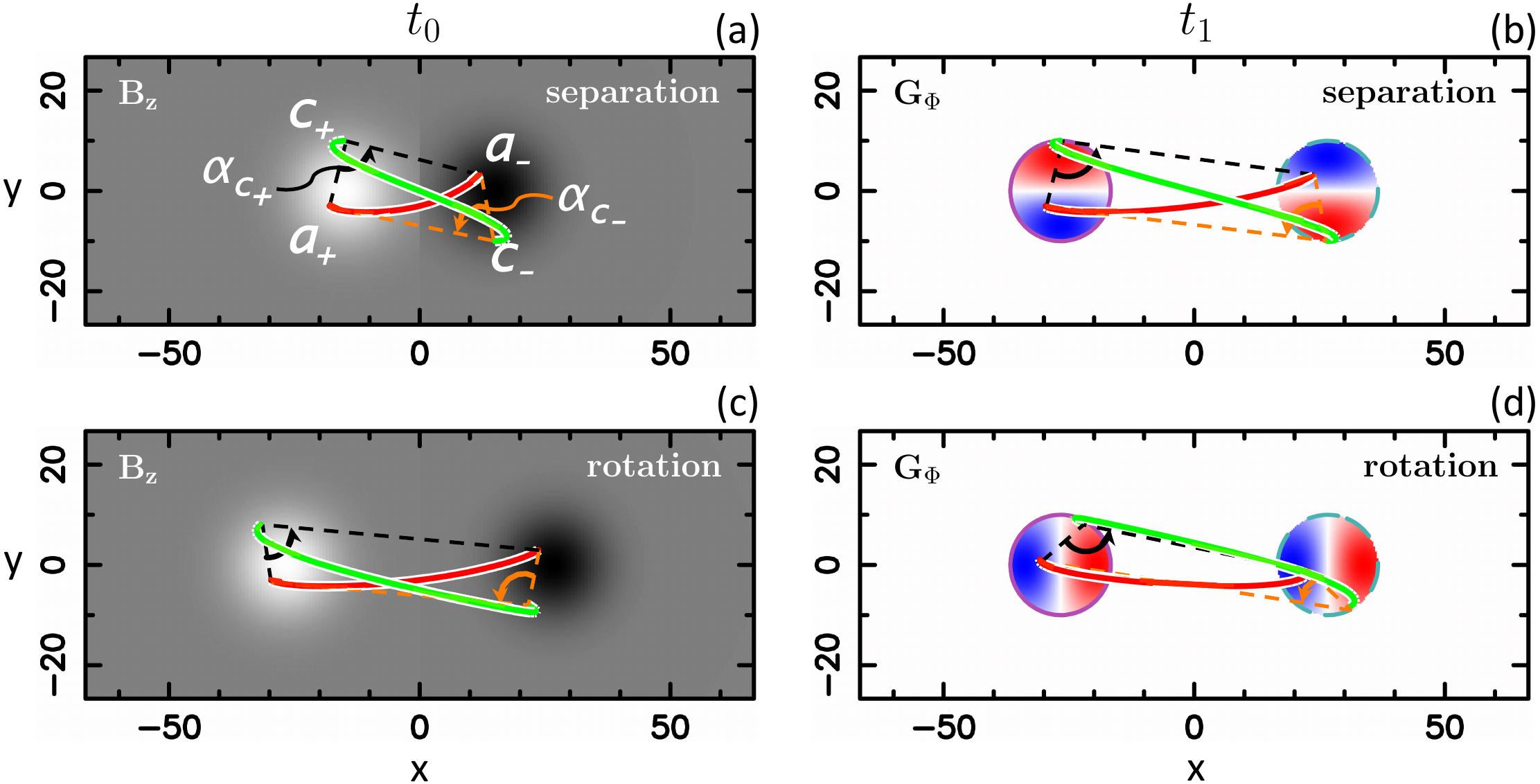}
              }
              \caption{Torus configuration for the twist $N=0.5$ illustrating the change in mutual magnetic helicity between two magnetic field lines (red and green lines) at two different times ($t_{1}>t_{0}$) and at $z=0$. 
    (a,b) Two separating magnetic polarities (as in \fig{Fig-Torus-sepx}).
    (c,d) Counter--rotating polarities (as in \fig{Fig-Torus-tworot}). 
    {\it Left column:} Normal component of the magnetic field at $t_{0}$. The values of the magnetic field are in the range $-1000 \ \textrm{G}$ (black) to $1000 \ \textrm{G}$ (white). 
    {\it Right column:} Distribution of $\gph$ at $t_{1}$. The angle between segments $c_{\pm}a_{\pm}$ and $c_{\pm}a_{\mp}$ represents the angle $\alpha_{c_{\pm}}$ of \eq{Eq-MutualH-angles}. The change in the relative orientation of the red and green magnetic field lines that leads to the change in $\alpha_{c_{\pm}}$ --- right panel compared with left panel --- is a clear evidence of mutual magnetic helicity changes of the two field lines between $t_{0}$ and $t_{1}$.
                      }
   \label{fig:Fig-Torus-mutualH}
   \end{figure}

\inlinecite{Demoulin06}, show that the total magnetic helicity 
flux in $\vol $ can be written as the summation of the mutual helicity 
of all pairs of elementary magnetic flux tubes contained in $\vol $, \ie 
the total magnetic helicity of the system can be 
rewritten as:
       \BE \label{eq:Eq-MutualH}
       	   H = \frac{1}{2\pi} \int_{\Phi_{\pm}} \int_{\Phi_{\pm}} \Hm_{a,c} \rmd \Phi_{a_{+}} \rmd \Phi_{c_{+}} \,,
       \EE
with $\Hm_{a,c}$, the mutual helicity between the two magnetic flux 
tubes {\it ``a''} and {\it ``c''}. Comparing the time derivative of \eq{Eq-MutualH}:
       \BE \label{eq:Eq-HFlux-vs-MutualH}
       	   \deriv{H}{t} = \frac{1}{2\pi} \int_{\Phi_{\pm}} \int_{\Phi_{\pm}} \deriv{\Hm_{a,c}}{t} \rmd \Phi_{a_{+}} \rmd \Phi_{c_{+}} \,,
       \EE
to \eqs{Eq-dH-Phi-doubleInt}{Eq-dhdens-Phi} implies:     
  
      \BA
      \deriv{\Hm_{a,c}}{t} \; &=& \; 
                     \deriv{ \theta (\xx _{c_+} -\xx _{a_-}) }{t}
                           + \deriv{ \theta (\xx _{c_-} -\xx _{a_+}) }{t}
                                               \nonumber  \\
       &&     \! \! \!
                     - \deriv{ \theta (\xx _{c_+} -\xx _{a_+}) }{t}
                         - \deriv{ \theta (\xx _{c_-} -\xx _{a_-}) }{t}          \,.
                     \label{eq:Eq-MutualH-Flux}
      \EA
By integrating \eq{Eq-MutualH-Flux} in time, they express the mutual 
helicity of two magnetic field lines {\it ``a''} and {\it ``c''} as a function of the 
angles between their photospheric footpoints. Using the convention that 
field line {\it ``c''} is above field line {\it ``a''}, they obtain: 
       \BE \label{eq:Eq-MutualH-angles}
       	  \Hm_{a,c} = \LLa_{a,\uvec{c}} = \frac{1}{2\pi} \left( \alpha_{c_{+}} + \alpha_{c_{-}} \right)  \,,
       \EE
where $\alpha_{c_{\pm}}$ is the angle between segments $c_{\pm}a_{\pm}$ 
and $c_{\pm}a_{\mp}$ and is defined in the interval $[-\pi,\pi]$ with the 
trigonometric convention (\fig{Fig-Torus-mutualH}a). 
The consequence is that any change in these angles will lead to a variation 
of mutual helicity and thus, a flux of magnetic helicity 
(\eqss{Eq-HFlux-vs-MutualH}{Eq-MutualH-angles}).

%:          Figure: Helicity flux density - Torus Mutual Helicity
%%%%%%%%%%%%%%%%%%%%%%%%%%%%%%%%%%%%%%%%%
%%%%%%%%%%%%%%%%%% FIGURE 9 %%%%%%%%%%%%%%%%%
%%%%%%%%%%%%%%%%%%%%%%%%%%%%%%%%%%%%%%%%%
%%%%%%%%%%%%%%%%%%%%%%%%%%%%%%%%%%%%%%%%% 

Let us consider the two magnetic field lines {\it ``a''} and {\it ``c''} starting 
at $a_{+}$ and $c_{+}$ and ending at $a_{-}$ and $c_{-}$, respectively 
represented by the red and green lines in \fig{Fig-Torus-mutualH} 
of the torus for $N=0.5$. As the two polarities of the torus separate away 
from each other in the $x$-direction, the $y$-coordinate of all four footpoints 
remains unchanged. Hence, the orientation of the segments $c_{\pm}a_{\pm}$ 
remains also the same, and only segments $c_{\pm}a_{\mp}$ change of orientation. 
In particular, the geometry implies that $\alpha_{c_{\pm}}$ increases 
as the polarities separate in the case shown in \fig{Fig-Torus-mutualH} for $N=0.5$. 
Therefore, the separation induces a positive variation of 
mutual helicity of {\it ``a''} and {\it ``c''}.

More generally, there is always an increase of mutual helicity of the magnetic 
field line {\it ``c''} in \figpnls{Fig-Torus-mutualH}{a}{b} with any other magnetic field line 
{\it ``a''} inside the polarities: as the polarities separate, there is always an increase 
of $\alpha_{c_\pm}$ for any {\it ``a''} within the polarities. 
This results in a net positive change of mutual helicity for {\it ``c''} with regard 
to all the other {\it ``a''}, and thus, a net positive helicity flux $\gph$ at the footpoint 
location of {\it ``c''}.

A more precise geometrical analysis --- \ie using the general definition of 
$\Hm_{a,c}$ in Equation (32) of \inlinecite{Demoulin06} 
--- reveals that, for the magnetic field 
lines footpoints of the $y>0$ (\resp $y<0$) part of the positive polarity, there 
is a net positive (\resp negative) variation of mutual helicity with a magnitude 
decreasing with $y$ (\resp increasing with $-y$).

For $N$ close to $0 \ [1]$ (\ie {\it modulo} $1$) turn, there is a similar behavior, 
except that the origin $O=(0,0)$ is no longer a center of symmetry for the magnetic 
footpoints. In particular, at these values of N, we find that, for the most 
externe (\resp interne) magnetic field lines, there is a net positive 
(\resp negative) variation of mutual helicity leading to a net positive 
(\resp negative) helicity flux (\fig{Fig-Torus-sepx}). However, the helicity 
flux at these $N$ values is typically ten times smaller when $N=0.015 \ [1]$ than 
for the $N=0.5 \ [1]$ case. As $N$ gets closer to $0.5 \ [1]$ turn, the magnetic 
field lines are more twisted and they share more mutual helicity, \ie the angles 
between the footpoints of two field lines are larger. The consequence 
is that, the change in the angles between footpoints, \ie in their mutual helicity, 
will be higher as the two polarities separate, tending towards the helicity flux 
distribution of \fig{Fig-Torus-sepx} as $N$ gets closer to $0.5 \ [1]$ turn.

Therefore, the signal in the $\gph$ maps of \fig{Fig-Torus-sepx} is due to 
a variation of mutual helicity between magnetic field lines as the two 
polarities separate, and thus, is a real signal.

\subsection{One Polarity Rigidly Rotating Around the Other} \label{sec:S-Torus-onerot}

The flux transport velocity field is given by \eq{Eq-FTV-onerot}. 
As in \sect{S-Charges-onerot}, the helicity flux density distribution 
computed using $\gth$ only presents helicity flux in the positive polarity 
as in \fig{Fig-Charges-onerot}a. The computation of helicity injection using $\gph$ 
removes this problem and the associated flux is typically twice 
smaller than in $\gth$, but present on both polarities independently 
of N value (comparable to \fig{Fig-Charges-onerot}b).

\subsection{Two Counter--rotating Polarities} \label{sec:S-Torus-tworot}

As pointed out in \sect{S-Charges-tworot}, for slow enough motions of 
the flux transport velocity field (given by \eq{Eq-FTV-tworot}), this model 
is equivalent to a cylinder rotating around its axis. The only difference is 
that, now, the cylinder has twisted magnetic field lines. The presence of 
twisted field lines, though, does not change the fact that no helicity is 
globally injected to the system. 

\begin{figure}[h]
   \centerline{\includegraphics[width=0.99\textwidth,clip=]{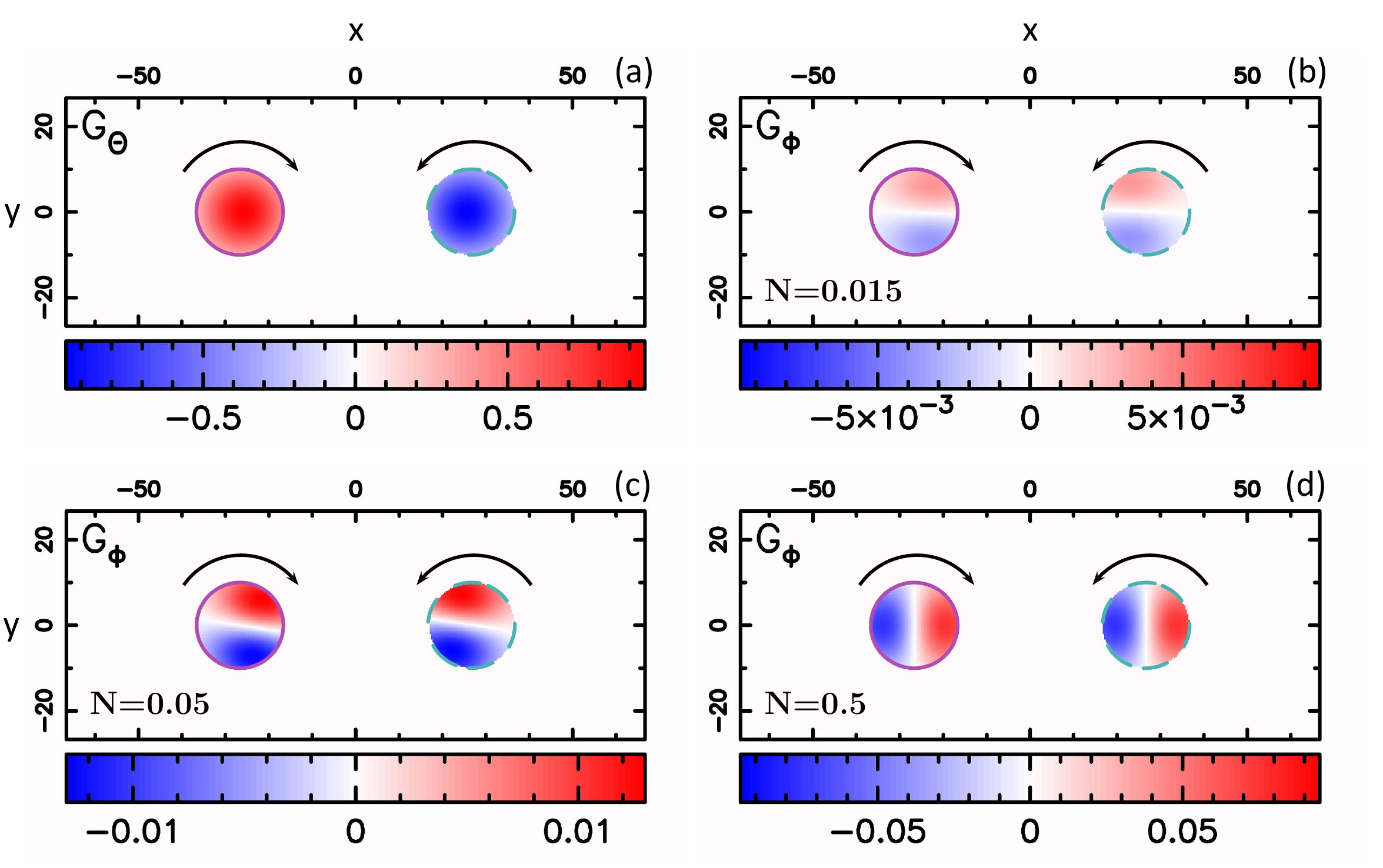}
              }
              \caption{$\gth$ and $\gph$ maps for the torus magnetic field configuration with two counter--rotating magnetic polarities. (a) $\gth$ map. (b,c,d) $\gph$ maps for $N=0.015$, $0.05$ and $0.5$ respectively. The drawing convention is the same as in \fig{Fig-Charges-sepx} (notice the different color scales).
                      }
   \label{fig:Fig-Torus-tworot}
   \end{figure}

The instantaneous footpoint injections displayed by the $\gth$ map, 
\fig{Fig-Torus-tworot}a, are similar to those in \fig{Fig-Charges-tworot}a. 
They would be meaningful for very fast motions.

%:          Figure: Helicity flux density - Torus counter--rotation
%%%%%%%%%%%%%%%%%%%%%%%%%%%%%%%%%%%%%%%%%
%%%%%%%%%%%%%%%%%% FIGURE 10 %%%%%%%%%%%%%%%%%
%%%%%%%%%%%%%%%%%%%%%%%%%%%%%%%%%%%%%%%%%
%%%%%%%%%%%%%%%%%%%%%%%%%%%%%%%%%%%%%%%%%  

Considering $\gph$ maps, the distribution changes significantly 
depending on the degree of twist (\fig{Fig-Torus-tworot}). 
While the global injection stays null, $\gph$ maps reveal subtle effects 
of mutual helicity variation between the twisted lines within the flux rope. 
Because of the twist, the magnetic field lines of the torus share mutual 
helicity between each other as the flux rope globally rotates around its axis. 
In a way similar to what has been discussed in \sect{S-Torus-sepx},  as the 
two polarities counter--rotate, the relative orientation of the magnetic field 
lines within the flux rope changes: \cf \figpnls{Fig-Torus-mutualH}{c}{d}. 
The magnitude of this variation increases with the number of turns, $N$, 
of the magnetic field lines around the torus axis.
This induces a net change of mutual helicity between the magnetic field 
lines revealed by the $\gph$ maps of \fig{Fig-Torus-tworot}. This process 
is completely hidden by the $\gth$ maps which are completely independent 
of the twist amount.

The use of $\gph$ is here crucial to understand the variation of 
mutual magnetic helicity driven by photospheric footpoint motions.

%%%%%%% U N I F O R M %%%%%%%
\section{Results for Extrapolated Magnetic Fields} \label{sec:S-Uniform}

In observational studies, extrapolations of the magnetic field will be used 
to infer the connectivity. Hence, in our tests, we consider extrapolated 
magnetic fields in order to study the influence of using them on 
the precision of the connectivity and thus, of $\gph$.

In this section, we consider two uniform opposite magnetic polarities 
with $B_{0}=1000 \ G$ in the positive polarity and $-1000 \ G$ in the negative 
polarity. Three linear force--free fields are considered in all our three flux 
transport velocity fields investigated: a potential field, and two linear force--free 
fields with a force--free parameter equal to $\alpha_{1} = 10^{-3}$ 
and $\alpha_{2}=5.6 \times 10^{-3} \ \textrm{Mm}^{-1}$ (\figpnls{Fig-Bfields}{c}{e}) which 
are typical values derived from observations (see \eg \opencite{Pevtsov95}; 
\opencite{Longcope03}; \opencite{Green02b}; \opencite{Chandra10}).

\subsection{Two Separating Magnetic Polarities} \label{sec:S-Uniform-sepx}
	
In this section, we study the distribution of helicity flux density for the 
two separating magnetic polarities case (see \eq{Eq-FTV-sepx}). 

\begin{figure}[h]
   \centerline{\includegraphics[width=0.99\textwidth,clip=]{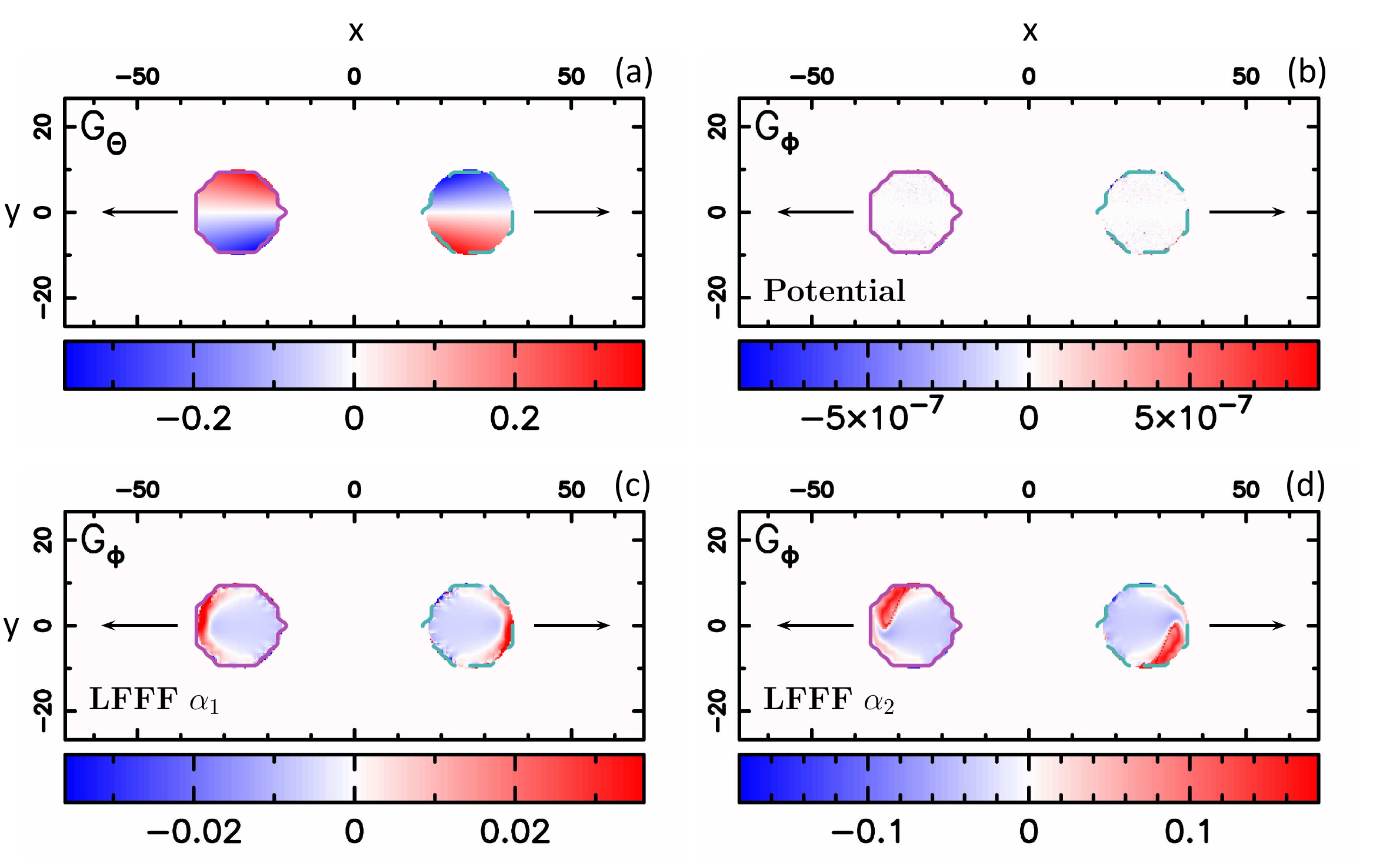}
              }
              \caption{$\gth$ and $\gph$ maps for the extrapolated magnetic field configurations with two separating magnetic polarities. (a) $\gth$ map. (b,c,d) $\gph$ maps for the linear force--free fields $\alpha=0$ (potential), $\alpha_{1}=10^{-3}$ and $\alpha_{2}=5.6 \times 10^{-3} \ \textrm{Mm}^{-1}$. The drawing convention is the same as in \fig{Fig-Charges-sepx} (notice the different color scales).
                      }
   \label{fig:Fig-Uniform-sepx}
   \end{figure}

First, let us consider the potential field ($\alpha=0$, \fig{Fig-Uniform-sepx}b). 
With this magnetic field configuration, the model is analogous to the one considered 
\sect{S-two-source-charges}. Consequently, as in \sect{S-Charges-sepx}, we expect 
no signal in the $\gph$ map as no helicity is injected to the system. This is 
well shown in \fig{Fig-Uniform-sepx}b where the 
helicity flux signal is indeed null to the numerical errors.
	
Let us now consider the two linear force--free magnetic field configurations.  
These configurations are analogous to the torus one (for $N \neq 0$) in the 
sense that magnetic field lines now have non--null mutual helicity. Therefore, 
as the two polarities separate away from each other, the angles between 
magnetic field lines footpoints change. This results in a variation of mutual 
helicity between field lines inducing a local flux of magnetic helicity as shown 
in \figpnls{Fig-Uniform-sepx}{c}{d}. However, the total helicity flux is 
indeed zero as expected (\sect{S-Torus-sepx}).

%:          Figure: Helicity flux density - Extrapolated separation
%%%%%%%%%%%%%%%%%%%%%%%%%%%%%%%%%%%%%%%%%
%%%%%%%%%%%%%%%%%% FIGURE 11 %%%%%%%%%%%%%%%%%
%%%%%%%%%%%%%%%%%%%%%%%%%%%%%%%%%%%%%%%%%
%%%%%%%%%%%%%%%%%%%%%%%%%%%%%%%%%%%%%%%%%  

Note also that, while the $\gth$ map exhibits, in each polarity, two regions of 
strong net opposite helicity flux (symmetric with regard to the $x$-axis), the 
$\gph$ maps present a diffuse (concentrated) region of negative (positive) flux
in the inner (most external) part of the system, respectively. 
In addition, the higher the linear force-free field constant--$\alpha$ is (in 
magnitude), the higher is the magnitude of the helicity flux signal in each 
polarity. These results are in agreement with the ones for the torus case and, 
again, demonstrate the limits of the $\gth$ proxy.

\subsection{One Polarity Rigidly Rotating Around the Other} \label{sec:S-Uniform-onerot}

In this section, the negative polarity rigidly rotates around the center of the positive 
polarity (\eq{Eq-FTV-onerot}).

Because $\gth$ does not take the magnetic field lines connectivity into account, 
it is not able to show that helicity is injected in both magnetic polarities. As in 
\sects{S-Charges-onerot}{S-Torus-onerot}, $\gph$ proxy displays the true distribution of helicity flux, 
which is positive in both positive and negative polarities and twice smaller than 
with $\gth$ in the positive polarity (comparable to \fig{Fig-Charges-onerot}).

\subsection{Two Counter--rotating Polarities} \label{sec:S-Uniform-tworot}

The two opposite magnetic polarities are counter--rotating around their own center 
(\eq{Eq-FTV-tworot}). We recall that this model is equivalent to the rotation of 
a cylinder around its axis, and hence, no helicity is globally injected to the 
system. 

\begin{figure}[h]
   \centerline{\includegraphics[width=0.99\textwidth,clip=]{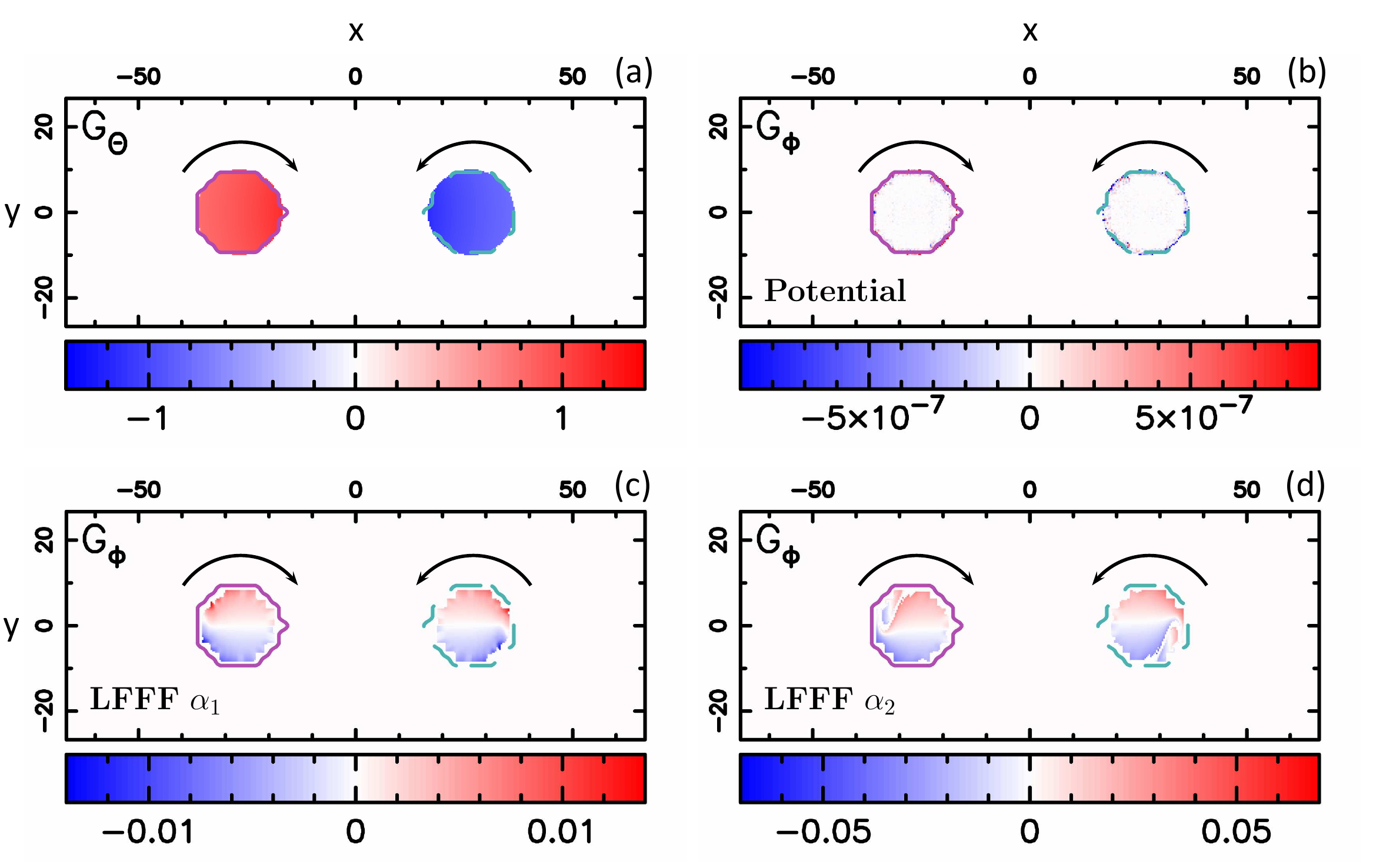}
              }
              \caption{$\gth$ and $\gph$ maps for the extrapolated magnetic field configurations with two counter--rotating magnetic polarities. (a) $\gth$ map. (b,c,d) $\gph$ maps for the linear force--free fields $\alpha=0$ (potential), $\alpha_{1}=10^{-3}$ and $\alpha_{2}=5.6 \times 10^{-3} \ \textrm{Mm}^{-1}$. The drawing convention is the same as in \fig{Fig-Charges-sepx} (notice the different color scales).
                      }
   \label{fig:Fig-Uniform-tworot}
   \end{figure}

For the same reasons as in \sect{S-Charges-tworot}, the $\gph$ map of the 
potential magnetic field case has zero values (to the numerical errors) 
everywhere.  As indicated by \fig{Fig-Uniform-tworot}, the $\gph$ maps for the two linear 
force--free fields ($\alpha \neq 0$) present non--zero helicity fluxes with both 
signs in both polarities. As for the torus case ($N \neq 0$), the signal in $\gph$ 
maps is real as there is a change of mutual helicity between magnetic field lines 
as the two polarities rotate.

%:          Figure: Helicity flux density - Extrapolated counter--rotation
%%%%%%%%%%%%%%%%%%%%%%%%%%%%%%%%%%%%%%%%%
%%%%%%%%%%%%%%%%%% FIGURE 12 %%%%%%%%%%%%%%%%%
%%%%%%%%%%%%%%%%%%%%%%%%%%%%%%%%%%%%%%%%%
%%%%%%%%%%%%%%%%%%%%%%%%%%%%%%%%%%%%%%%%%  

\subsection{Errors Estimation} \label{sec:S-Uniform-error}

As in \sect{S-Charges-error}, we estimate the computation errors due 
to the magnetic field lines integration as a function of the field lines 
integration parameter $n$. An analytical connectivity is available 
for the potential field since it has the same type of connectivity as the two 
magnetic charges (although with a different $\Bn$ distribution). Then, it 
is straightforward to compute the theoretical $G_{\Phi,\rm{th}}$ value for each flux 
transport velocity model and the errors.
 
\begin{figure}[h]
   \centerline{\includegraphics[width=0.59\textwidth,clip=]{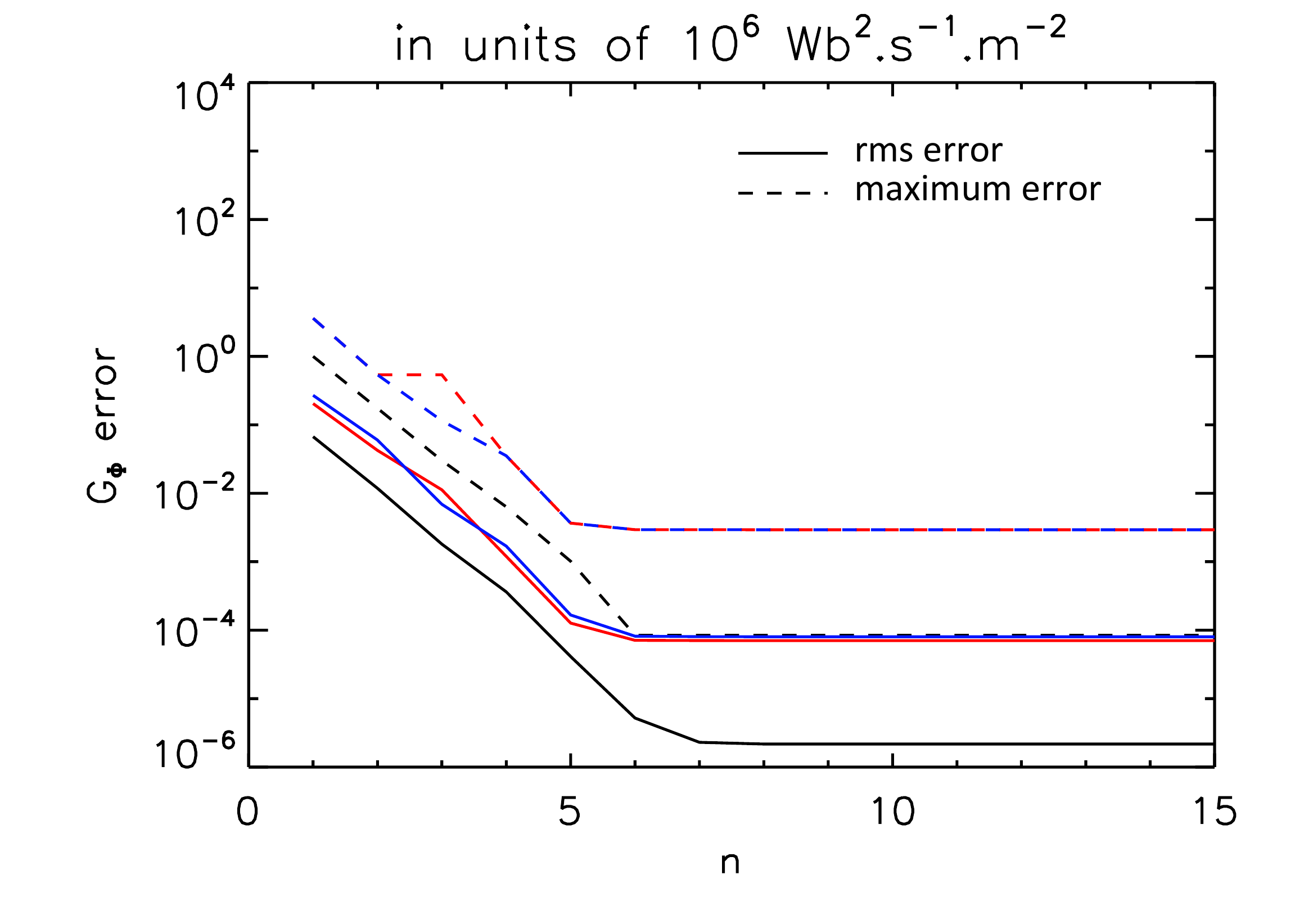}
              }
              \caption{Computed root mean square (solid lines) and maximum error (dashed lines) from $\gph$ maps as a function of the field line integration parameter $n$ defining the numerical integration precision ($=10^{-n}$). This plot is similar to \fig{Fig-Charges-error} but done for an extrapolated potential field. Three boundary flows are shown: two separating polarities (black), the negative polarity rigidly rotating around the positive one (red), and two counter--rotating magnetic polarities (blue).
                      }
   \label{fig:Fig-Uniform-error}
   \end{figure}

As expected, \fig{Fig-Uniform-error} shows that both $\gph$ rms and maximum error 
exponentially decrease as $n$ increases. However, it also shows a saturation 
of the errors to $10^{-6}-10^{-4}$ $\gth$ units for $n >6$. 
In particular, the rms on $\gph$ (black, red and blue solid lines) 
saturate at $0.2$, $7$ and $8 \times 10^{-5}$ $\gth$ units respectively.

The use of an extrapolated magnetic field implies that the magnetic field 
is discretized. Hence, at each step of the integration of magnetic field lines, 
the magnetic field is interpolated and not analytically computed. 
This affects the precision on the computation of $\Bn$ and $\gth$ at the second footpoint 
of each magnetic field line, \ie enhances the terms ${\rm{err}}_{\Bn}$ and ${\rm{err}}_{{\rm{interp}}}$ in 
\eq{Eq-guesstimated-error}. 
This is well illustrated in 
\fig{Fig-Uniform-error} where the choice of $n$ dominates the precision 
of $\gph$ only up to $n=6$. 
Even though the precision reached on 
$\gph$ is much less for $n > 6$ than in the potential analytical case 
(\sect{S-two-source-charges}), our method still allows us to compute the true 
distribution of helicity injection with a good accuracy.

%:          Figure: Extrapolated - Gphi errors estimation
%%%%%%%%%%%%%%%%%%%%%%%%%%%%%%%%%%%%%%%%%
%%%%%%%%%%%%%%%%%% FIGURE 13 %%%%%%%%%%%%%%%%%
%%%%%%%%%%%%%%%%%%%%%%%%%%%%%%%%%%%%%%%%%
%%%%%%%%%%%%%%%%%%%%%%%%%%%%%%%%%%%%%%%%%   

\section{Results for Magnetic Fields Containing Quasi-separatrix Layers} \label{sec:S-Hinj-in-QSLs}
	
In this section, we investigate the helicity injection at the $z=0$--plane in 
two quadrupolar magnetic field configurations with quasi-separatrix layers 
(QSLs), from the simulations of \inlinecite{Aulanier05}. Our goal is to study 
the quality of our method when strong connectivity gradients are present.

\subsection{QSLs} \label{sec:S-QSLs}

  %{\S}{\bf --- QSLs def} \\
	
QSLs are regions where the magnetic field lines connectivity changes 
continuously with very sharp gradients with the limit case of separatrices
when gradients are infinite \cite{Demoulin96a,Titov02}. 
Even in the cases with continuous connectivity changes, QSLs are 
preferential sites for current layers formation 
(see \inlinecite{Aulanier05} and references therein).

  %{\S}{\bf --- QSLs def in terms of N & Q} \\
  
The concept of QSLs has been intensively studied and developed in the 
last two decades (see review by \inlinecite{Demoulin06Review} and 
references therein) and observational data analyses have reported the 
presence of such topological structures in the solar atmosphere 
(\eg \opencite{Demoulin97}; \opencite{Mandrini97}, \citeyear{Mandrini06}; 
\opencite{Bagala00}; \opencite{Masson09}; \opencite{Baker09}; 
\opencite{Savcheva12}). 
QSLs are defined as regions where the squashing degree, $Q$, is much larger 
than $2$ \cite{Titov02}. If we 
consider an elementary flux tube --- within a QSL --- with one circular photospheric 
footpoint, then $Q$ is a measure of the squashing of the section of this elementary 
flux tube at the other photospheric footpoint. 
Configurations with QSLs are thus cases for which a 
connectivity--based helicity flux density is required to localize the true site(s) of 
helicity injection and, \eg , study its role in the trigger of eruptive events.

\subsection{Initial Magnetic Field Configurations and Flux Transport Velocities}	
 \label{sec:S-QSLs-Initial}		
  %{\S}{\bf --- Magnetic field configurations - Aulanier etal 05} \\

In the following, we consider the magnetic field configurations 
from the simulations of \inlinecite{Aulanier05} on the formation of current layers 
in QSLs (\fig{Fig-Bfields}f). The magnetic configurations are referred to as $\Phi=120^{\circ}$ and 
$\Phi=150^{\circ}$ where $\Phi$ describes the angle between the inner and outer 
dipoles. For each magnetic configuration, two flux transport velocity fields are 
considered: a nearly solid translation in the $y$-direction and a nearly solid 
rotation of the positive polarity of the inner dipole (see Sections~2,3 and Figure~5 of 
\inlinecite{Aulanier05} for further detailed informations on the setup). 
For simplicity, the positive and negative polarities of the inner or outer dipole
will be referred to as IP and IN, or OP and ON, respectively.

\subsection{Results with Twisting Motions} \label{sec:S-QSLs-twist}

In this model, the IP polarity nearly rigidly rotates counterclockwisely 
around its center. The flux transport velocity field is given by Equations (13) and (14) 
of \inlinecite{Aulanier05}. In terms of helicity 
flux density, we can analytically show that $\gth (\xx)=0$ when $\xx$ is not 
in IP and $\gth (\xx \ \textrm{in} \ \rm{IP})<0$. 
For $\gph$ on the other hand, because the twisting motion is 
applied to one part of the QSLs, we expect to see two regions of a twice 
smaller helicity flux: in IP and in the part of the QSLs connected to it.

%:          Figure: Helicity flux density - QSLs twist
%%%%%%%%%%%%%%%%%%%%%%%%%%%%%%%%%%%%%%%%%
%%%%%%%%%%%%%%%%%% FIGURE 14 %%%%%%%%%%%%%%%%%
  \begin{figure}[h]
   \centerline{\includegraphics[width=0.97\textwidth,clip=]{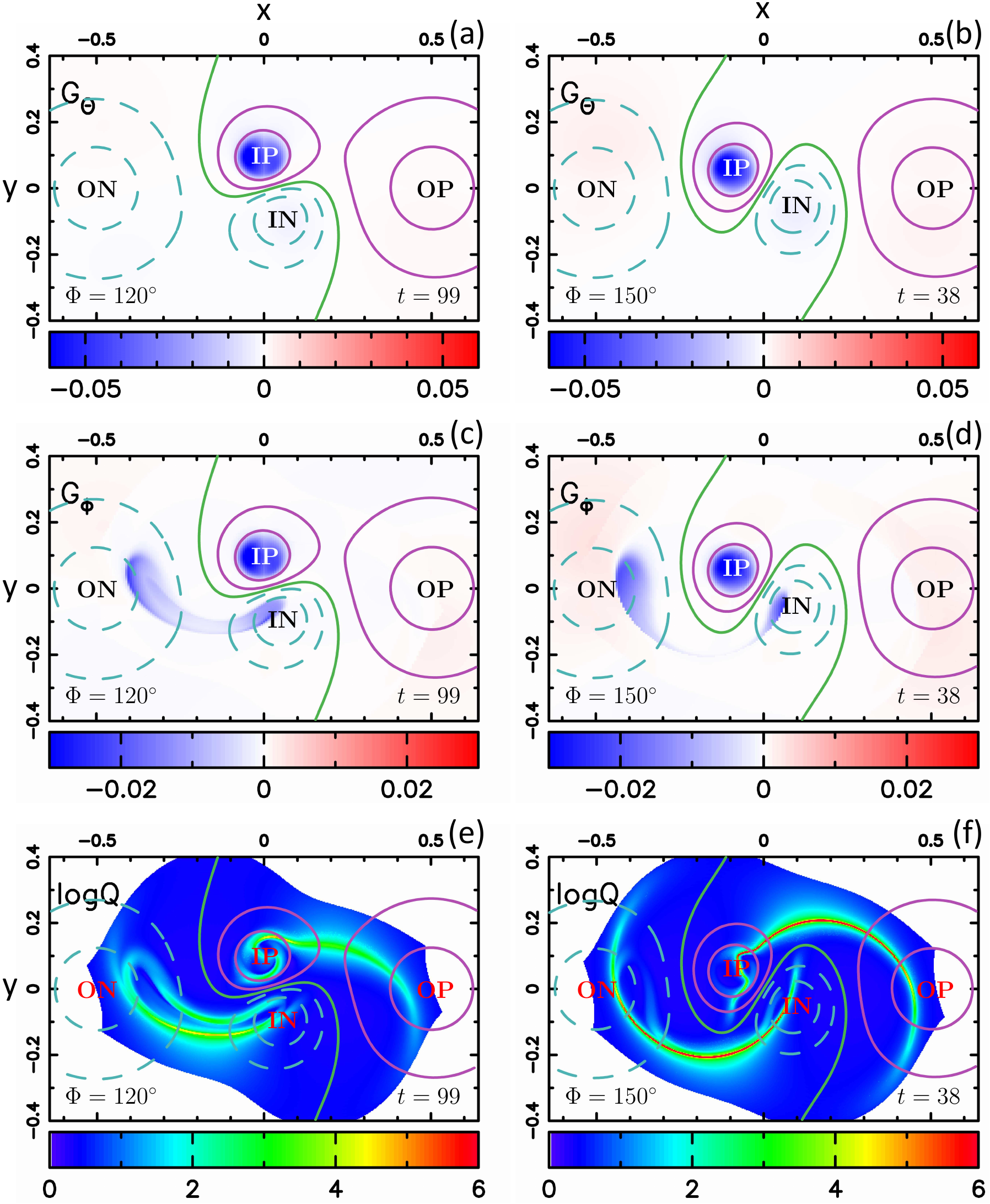}
              }
              \caption{Results of two MHD simulations with a nearly solid rotation of the inner magnetic polarity (IP) and no motion in other polarities. The angle between the inner and outer bipoles is $\Phi=120^{\circ}$ (left column) and $\Phi=150^{\circ}$ (right column) and the simulation time is $t=99$ and $t=38$ Alfv\'en times respectively. The panels show the photospheric distribution of: (a,b) $\gth$, (c,d) $\gph$, and (e,f) ${\rm{log}}_{10}Q$. The green line corresponds to the polarity inversion line, while the solid purple (\resp dashed cyan) are positive (\resp negative) isocontours of the magnetic field.  
                      }
   \label{fig:Fig-QSLs-twist}
   \end{figure} 
%%%%%%%%%%%%%%%%%%%%%%%%%%%%%%%%%%%%%%%%%
%%%%%%%%%%%%%%%%%%%%%%%%%%%%%%%%%%%%%%%%%  

\fig{Fig-QSLs-twist} displays the results of the $\gth$ (top row) and $\gph$ 
(middle row) computations for the $\Phi=120^{\circ}$ (left column) and 
$150^{\circ}$ (right column) configurations. 
As expected, $\gth$ maps present a negative helicity flux distributed only in IP. 

In the $\gph$ maps, two main distinct regions of negative helicity flux 
are present (\figpnls{Fig-QSLs-twist}{c}{d}).  The first region, IP, has a flux 
twice smaller than in $\gth$ as expected (notice the factor $1/2$ 
between $\gth$ and $\gph$ color scales). The 
second region corresponds to the QSL portion connected to IP. 
The helicity flux is more concentrated on the edges of the QSL in both 
ON and IN (see $\gph$ compared to ${\rm{log}}_{10}Q$ maps). This effect is 
even more important for the $\Phi=150^{\circ}$ configuration (\fig{Fig-QSLs-twist}d). 
This is due to the $|\Bn (\xx_{a_{-}})/\Bn (\xx_{a_{+}})|$ ratio which is 
much smaller than unity in between ON and IN. 
Hence, from \eq{Eq-Gphi}, it results that the helicity flux density is 
much weaker in the center of the QSL than at its edges. Because in the $\Phi=120^{\circ}$ 
configuration, IN and ON are closer to each other, the region of weak $\Bn$ is smaller and 
the helicity flux distribution appears less concentrated at the edges of the QSL than for the 
$\Phi=150^{\circ}$ configuration (\figpnls{Fig-QSLs-twist}{c}{d}).

Finally, we notice that, the positive helicity flux signal in $\gph$ maps 
(\figpnls{Fig-QSLs-twist}{c}{d}) is a remnant spurious signal already present 
in $\gth$ maps, as the velocity field is not numerically limited to IP.

%\newpage
	
\subsection{Results with Translational Motions} \label{sec:S-QSLs-trans}

%:          Figure: Helicity flux density - QSLs translation
%%%%%%%%%%%%%%%%%%%%%%%%%%%%%%%%%%%%%%%%%
%%%%%%%%%%%%%%%%%% FIGURE 15 %%%%%%%%%%%%%%%%%
  \begin{figure}[h]
   \centerline{\includegraphics[width=0.97\textwidth,clip=]{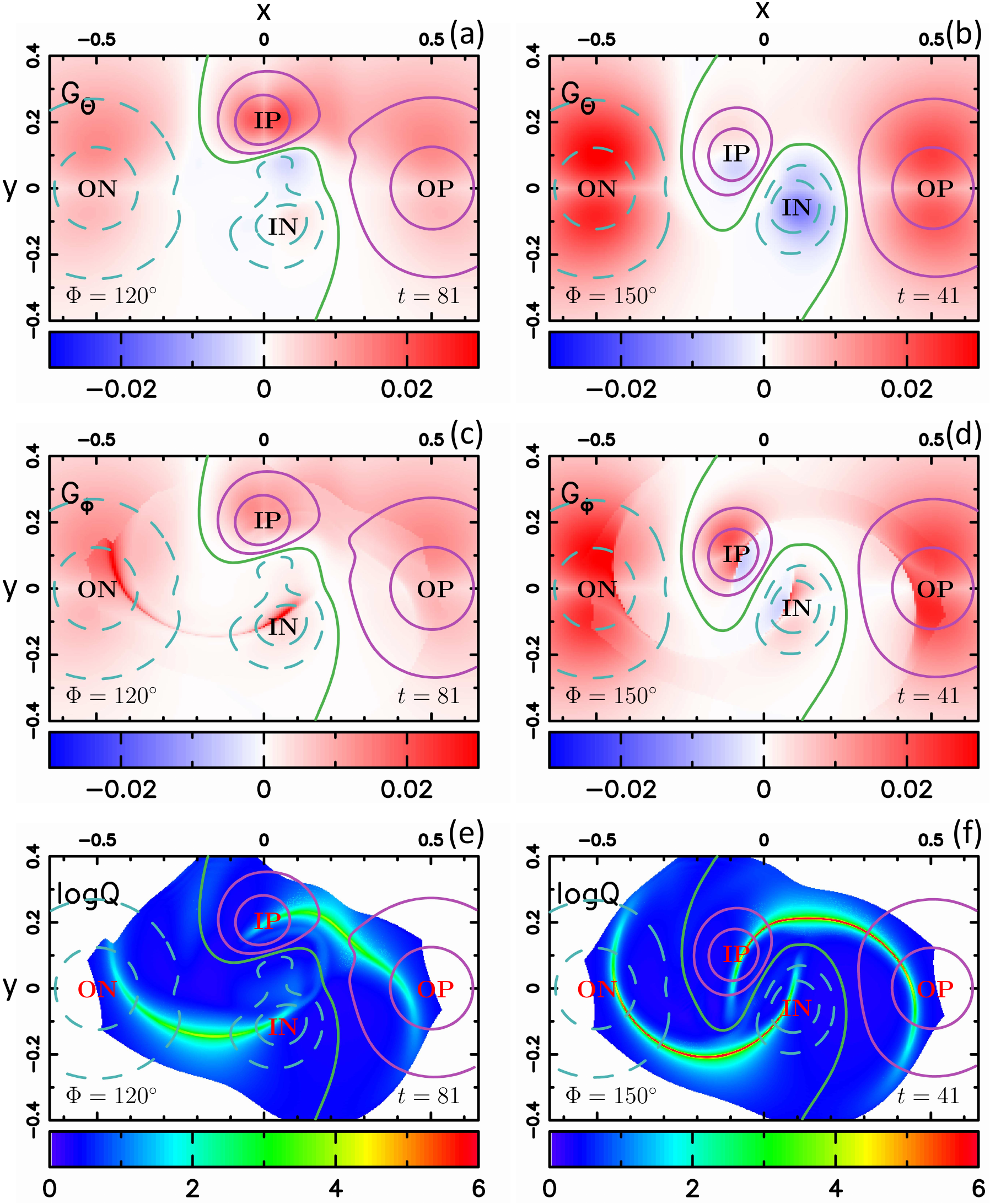}
              }
              \caption{Results of two MHD simulations with a nearly solid translation of the inner magnetic polarity (IP) and nearly no motion in other polarities. The angle between the inner and outer bipoles at $t=0$ is $\Phi=120^{\circ}$ (left column) and $\Phi=150^{\circ}$ (right column) and the simulation time is $t=81$ and $t=41$ Alfv\'en times respectively. The panels show the photospheric distribution of: (a,b) $\gth$, (c,d) $\gph$, and (e,f) ${\rm{log}}_{10}Q$. The drawing convention is the same as in \fig{Fig-QSLs-twist}.
                      }
   \label{fig:Fig-QSLs-trans}
   \end{figure} 
%%%%%%%%%%%%%%%%%%%%%%%%%%%%%%%%%%%%%%%%%
%%%%%%%%%%%%%%%%%%%%%%%%%%%%%%%%%%%%%%%%%  

Let us now consider the nearly translational motion of IP in the $y$-direction 
--- given by Equation (12) of \inlinecite{Aulanier05} --- 
which leads to a global shearing of the configuration (\fig{Fig-QSLs-trans}). 
We note that a small part of IN is also affected by the numerical setup 
(see the deformation of the isocontours of IN compared with the twisting case). 

First, let us consider the 
$\Phi=120^{\circ}$ case. The $\gth$ map (\fig{Fig-QSLs-trans}a) 
shows a rather diffuse positive helicity 
injection in the outer dipole, a slightly more concentrated positive flux in IP, and 
a quasi--null flux around the IN except 
two spots of small negative and positive flux. 
From a theoretical point of view, $\gth$ can be divided into three contributions: 
the motion of IP with regard to OP, ON, and IN. 
Figure 5 of \inlinecite{Pariat06} and 
Figure 6 of \inlinecite{Pariat07c} can be used to infer the resulting sign of 
helicity injection of these three contributions.
The motion of IP with regard to the outer dipole is a shearing 
motion and injects positive helicity. The motion of IP with regard to the IN 
injects some negative helicity flux. However, IP and IN 
are almost aligned with the direction of the translational motion. Hence, the associated 
shearing is much weaker than that of IP with regard to the outer dipole. The 
resulting helicity flux is therefore much weaker. The summation of these three 
contributions explains the observed features in the $\gth$ map. 
The associated $\gph$ map only exhibits positive helicity flux 
signal. In particular, it shows that helicity is also injected in the QSL connected 
to IP (see \figpnls{Fig-QSLs-trans}{c}{e}), with a stronger flux at the edges of the QSL 
(for the same reason as in \sect{S-QSLs-twist}). It also presents 
some weak positive helicity injection that allows to slightly distinguish the 
QSL in the positive polarity.

The flux transport velocity field of the $\Phi=150^{\circ}$ configuration implies a 
global shearing as for the $\Phi=120^{\circ}$ case. The difference is that now, 
the inner dipole is more aligned with the outer dipole. Hence, as IP moves, 
the shearing of the inner dipole is more negative than for $\Phi=120^{\circ}$. Therefore, the total 
helicity flux distribution in IP is a sum of positive --- from shearing with ON and 
OP --- and negative --- from shearing with IN --- fluxes. This explains the resulting 
$\gth$ map (\fig{Fig-QSLs-trans}b). On the other hand, $\gph$ map presents 
mainly positive helicity injection (\fig{Fig-QSLs-trans}d). In particular, a strong positive 
helicity injection is present in both external parts of the QSLs (\figpnls{Fig-QSLs-trans}{d}{f}). 
In the inner part, where IP and IN are magnetically connected, negative helicity 
injection is present. Such a result is actually expected since the white magnetic 
field lines of \fig{Fig-Bfields}f are sheared oppositely to 
the other magnetic field lines as IP translates towards the $y > 0$. 

Note that, for the $\gph$ maps considered in both nearly rigid rotation and translation 
of IP, for $43 \%$ of the photospheric mesh footpoints, the field lines integration did 
not lead to a second footpoint on the $z=0$--plane (open--like magnetic field 
lines reaching the boundary of the mesh). Therefore, for these $43 \%$ photospheric 
footpoints --- localized at the white regions of the squashing degree maps --- the helicity flux 
density $\gph$ is set equal to $\gth$. This is observable in the $\gph$ map of the 
translation model (\fig{Fig-QSLs-trans}c), where we can identify an abrupt change 
of $\gph$ at the limit of white and blue regions of \fig{Fig-QSLs-trans}e. In order to compute 
$\gph$ in a more extended part of the outer polarities, a larger numerical box is needed in 
the numerical simulations.

\subsection{Errors Estimation} \label{sec:S-QSLs-error}
	
For both $\Phi$ configurations and both flux transport velocity 
fields, we compute the total helicity flux computed from  $\gth$ and 
$\gph$ for all simulations output files as a function of time. We then 
estimate the errors on the total helicity flux computed from $\gph$ 
compared to $\gth$ by computing the rms of the difference of total fluxes. 
We find that the rms is $\approx 10^{-7}-10^{-6}$ 
while the values of this total flux are $\approx10^{-4}-10^{-3}$. Although the total 
helicity fluxes computed from both helicity flux densities are mathematically 
strictly equal, numerically we find tiny differences due to the precision in 
$\gph$ computation (see \eq{Eq-guesstimated-error}). However, they are 
typically $\approx 10^{3}$ times smaller than the typical values of the total helicity 
flux.

\section{Conclusions} \label{sec:S-Conclusion}

  %{\S}{\bf --- Aim of the paper} \\

In this paper, we focus on the study of the flux of magnetic helicity through 
the photosphere. As magnetic helicity is a global 
3D quantity, a density of magnetic helicity flux is only meaningful when 
defined by elementary magnetic flux tubes \cite{Pariat05}. 

Our aim is to present the first 
implementation of a method that computes the helicity flux density at the 
photosphere by taking into account the magnetic field connectivity. 
In order to test our method and use it in future observational studies, we have 
performed a comparative analysis of the distribution of helicity injection 
at the photosphere using two proxies of helicity flux density: $\gth$ and 
$\gph$. We have analyzed their properties on simplified solar 
configurations considering analytical, extrapolated magnetic 
fields and fields from numerical simulations. 

We find that, while the total helicity flux remains the same 
using $\gth$ or $\gph$, the distribution of helicity flux, however, can be 
significantly different. Using several test--cases, we confirm that 
$\gth$ does not always reveal the true distribution of helicity flux while 
$\gph$ properly localizes the true site(s) of helicity injection. 
In particular, we show that $\gth$ can hide subtle variation of mutual 
helicity between neighboring field lines in a flux tube (\cf \sects{S-Torus}{S-Uniform}).
We also analyze the effect of strong connectivity gradients on the helicity 
distribution in systems containing QSLs.
The error estimations highlight that our method of computing the 
field lines connectivity is very accurate using analytical and extrapolated 
magnetic fields as well as for magnetic fields from numerical simulations.

We finally discuss that some differences between $\gth$ and $\gph$ 
maps are related to the underlying assumptions of field lines connectivity. 
$\gth$ provides the locally injected helicity flux when the injection timescale 
is much shorter than the transit Alfv\'en time between field line footpoints. 
$\gph$ assumes that both magnetic field line footpoints are ``aware'' of 
the evolution of one another, hence, that the injection timescale is longer 
than the transit Alfv\'en time which is typically the case in most solar 
applications.

The use of the method that we have presented here will be quite useful 
when applied to actual observed ARs. For ARs with helicity 
flux density maps of uniform sign, while not changing the uniform 
character of the helicity injection, $\gph$ will enable to more precisely 
localize the regions where magnetic helicity is injected and accumulated. 
For ARs displaying mixed signs of helicity in $\gth$ maps 
\cite{Chandra10,Romano11a,Romano11b,Jing12}, $\gph$ will 
permit to remove the spurious mixed signal, displaying the true helicity 
flux distribution. It may result in a more complex and subtle 
injection of helicity, revealing mutual helicity changes between 
magnetic flux tubes, as in some examples presented in this study. 
$\gph$ will allow to more strictly determine which ARs present 
injection of opposite sign of magnetic helicity and relate this pattern 
to their eruptivity (\eg \opencite{Romano11b}). The $\gph$ maps 
will enable to observationally test the theoretical hypothesis that
more energy is eventually released when magnetic helicity annihilation 
occurs \cite{Kusano95,Linton01}. 
They will also allow to observationally test models based on magnetic helicity 
cancellation (\opencite{Kusano02}, \citeyear{Kusano04}).

Overall, $\gph$ will enable us to truthfully track the injection of 
helicity into the solar corona, helping us to better understand the role of 
magnetic helicity in solar activity.

%%%%%%%%%%%%%%%%%%%%%%%%%%%%%%%%%%%%%%%%%%%%%%%%%%%%%%%%%%%%%%%%%%%%%%%%%%%
\begin{acks}
 The authors thank A. Canou for providing the potential and linear force--free fields 
 computed with the XTRAPOL numerical code developed by T. Amari and 
 supported by the Centre National d'Etudes Spatiales \& the Ecole Polytechnique. 
 The authors thank the referee for helpful comments that improved the clarity of the 
 paper.
\end{acks}

%%%%%%%%%% Appendix %%%%%%%%%%
	
%	\newpage
	\appendix
	
\section{Analytical Solutions for $\gth$} \label{app:App-solutions}
	
In the following, we use the same notation as defined in \sect{S-methodo}. 
The flux transport velocity fields are given by \eqss{Eq-FTV-sepx}{Eq-FTV-tworot}. 
The total magnetic flux in the positive (\resp negative) polarity is called $\Phi_{+}$ 
(\resp $-\Phi_{-}$, with $\Phi_{\pm}>0$). For generality purpose, $\Phi_{-}$ can be different from 
$\Phi_{+}$, and no specific assumption is made concerning the magnetic field 
configuration.

\subsection{Two Separating Magnetic Polarities} \label{app:App-sepx}

We consider two opposite magnetic polarities separating in the 
$x$-direction at constant speed and without any rotation. The flux 
transport velocity field is given by \eq{Eq-FTV-sepx}. Because 
the velocity field is constant in each polarity, the terms of \eq{Eq-Gtheta} 
associated to $(M,M')$ in the same polarity are zero as a 
consequence of $\uu^{\prime}=\uu$.

In this case, we have $\uu-\uu'=\mp 2U_{0} \vec{e}_{x}$ when $\pm \Bn (M)>0$ 
and $\mp \Bn (M')>0$, which leads to:
      \BE
	  \vec{M'}\vec{M} \times (\uu-\uu') \vert_{n}= \pm 2U_{0} \vec{M'}\vec{M} \cdot  \vec{e}_{y} \,.
      \EE	
Thus, \eq{Eq-Gtheta} leads to:
      \BE
	\gth (M(\xx))=\mp \frac{U_{0} \Bn}{\pi} \left(\int_{M' \ \textrm{in} \ P_{\mp}} \Bn' 
	                   \frac{\vec{M'}\vec{M}}{|\vec{M'}\vec{M}|^{2}} \rmd \surf' \right) \cdot \vec{e}_{y} \,, 
	                   \ \textrm{for} \ \pm \Bn (M)>0 \,.
      \EE
This integral can be computed by analogy to the electric field created 
by a 2D distribution of charge, $\sigma (M)= \Bn'(M)$, of an infinite 
cylinder of radius $R$ (of vertical axis crossing the $z=0$ plane at 
point $O_{\mp}$), using Gauss theorem, \ie:
     \BE      \label{eq:Eq-App-solGauss-sepx}
	\int_{M' \ \textrm{in} \ P_{\mp}} \Bn'  \frac{\vec{M'}\vec{M}}{|\vec{M'}\vec{M}|^{2}} \rmd \surf'
	= \mp \frac{\vec{O}_{\mp} \vec{M}}{|\vec{O}_{\mp} \vec{M}|^{2}}\ \Phi_{\mp}    \,.
     \EE
Hence, we find that the helicity flux density is given by \eq{Eq-Gth-sepx}.

\subsection{One Polarity Rigidly Rotating Around the Other} \label{app:App-onerot}

In this model, the negative polarity rigidly rotates around the positive 
one. The velocity field is given by \eq{Eq-FTV-onerot}. There are four cases to consider.

\hspace{1cm} $\cdot$ c1. If $\Bn (M)>0$ and $\Bn (M')>0$, then $\uu-\uu'=0$ and the associated 
term of \eq{Eq-Gtheta} is null.

\hspace{1cm} $\cdot$ c2. If $\Bn (M)>0$ and $\Bn (M')<0$, 
then $\uu-\uu'= - \Omega \vec{e}_{z} \times {\vec O}_{+} \vec{M'}$ and:
     \BE
	\vec{M'}\vec{M} \times (\uu-\uu') \vert_{n}
	 = - (\vec{M'}\vec{M} \cdot \vec{O}_{+}\vec{M'}) \Omega  \,.
    \EE		
The helicity flux density is then (using $ \vec{O}_{+}\vec{M'} =  \vec{O}_{+}\vec{M} -  \vec{M'}\vec{M}$):
    \BA
	\gth (M(\xx)) &=& \frac{\Omega \Bn}{2\pi} 
	                 \int_{M' \ \textrm{in} \ P_{-}} \Bn' 
	                     \left( \frac{\vec{O}_{+}\vec{M} \cdot\vec{M'}\vec{M}}
	                          {|\vec{M'}\vec{M}|^{2}} -1 \right) \rmd \surf'  \nonumber \\ 
           \nonumber \\	                          
	                      &=&  \frac{\Omega \Bn}{2\pi} 
	                 \left( \vec{O}_{+}\vec{M} \cdot \int_{M' \ \textrm{in} \ P_{-}} \Bn' 
	                      \frac{\vec{M'}\vec{M}}
	                          {|\vec{M'}\vec{M}|^{2}} \rmd \surf' \ + \ \Phi_{-}   \right) \nonumber \\
           \nonumber \\	
	                      &=&  \frac{\Omega \Bn}{2\pi} 
	                 \left( - \ \frac{\vec{O}_{+}\vec{M} \cdot \vec{O}_{-}\vec{M}}
	                          {|\vec{O}_{-}\vec{M}|^{2}} \Phi_{-} \ + \ \Phi_{-}   \right) \,,  \ \textrm{for} \ \Bn (M)>0 \,,
    \EA
which, by regrouping terms, leads to \eq{Eq-Gth-onerot}.

\hspace{1cm} $\cdot$ c3. If $\Bn (M)<0$ and $\Bn (M')>0$, then 
$\uu-\uu'= \Omega \vec{e}_{z} \times {\vec O}_{+} \vec{M}$ and:
     \BE
	\vec{M'}\vec{M} \times (\uu-\uu') \vert_{n}=(\vec{M'}\vec{M} \cdot \vec{O}_{+}\vec{M}) \Omega  \,,
    \EE		
which, using \eq{Eq-App-solGauss-sepx} leads to:
    \BE 
	\gth (M(\xx)) = - \frac{\Omega \Bn}{2\pi}\ \Phi_{+}  \,,
	                  \ \textrm{for} \ \left(\Bn (M)<0, \Bn (M')>0 \right)  \,.
    \EE			

\hspace{1cm} $\cdot$ c4. If $\Bn (M)<0$ and $\Bn (M')<0$, $\uu-\uu'= \Omega \vec{e}_{z} \times {\vec M'}\vec{M}$ and:
    \BE
	\vec{M'}\vec{M} \times (\uu-\uu') \vert_{n}= |\vec{M'}\vec{M}|^{2} \Omega   \,,
    \EE
leading to:
    \BE
	\gth (M(\xx)) = \frac{\Omega \Bn}{2\pi}\ \Phi_{-}    \,,
	                  \ \textrm{for} \ \left(\Bn (M)<0, \Bn (M')<0 \right)  \,.
    \EE		
Then the total helicity flux density within the region $\Bn (M)<0$ is therefore:
    \BE
	\gth (M(\xx)) = - \frac{\Omega \Bn}{2\pi}\ (\Phi_{+} - \Phi_{-})    \,,
	                   \ \textrm{for} \ \Bn (M)<0 \,.
    \EE
Note that, in the particular case of two magnetic--flux balanced polarities, 
$\Phi_{+}=\Phi_{-}$ and $\gth=0$ for $\Bn (M)<0$.

\subsection{Two Counter--rotating Magnetic Polarities} \label{app:App-tworot}

In this model, the positive polarity rotates clockwise around its 
center, while the negative rotates counterclockwise around its 
center. The velocity field is given by \eq{Eq-FTV-tworot}. 
There are four cases to consider, which, by symmetry, reduce to two cases.

\hspace{1cm} $\cdot$ c1. If $\Bn (M)>0$ and $\Bn (M')>0$, we have 
$\uu-\uu'=-\Omega \vec{e}_{z} \times \vec{M'}\vec{M} $, which leads to:
     \BE
	\vec{M'}\vec{M} \times (\uu-\uu') \vert_{n}= -|\vec{M'}\vec{M}|^{2} \Omega    \,,
     \EE
giving:
     \BE     \label{eq:Eq-App-solGauss-tworot-11}
	\gth (M(\xx)) = \frac{\Omega \Bn }{2\pi}\ \Phi_{+}     \,,
	                   \ \textrm{for} \ \left(\Bn (M)>0, \Bn (M')>0 \right)  \,.
     \EE	

\hspace{1cm} $\cdot$ c2. If $\Bn (M)>0$ and $\Bn (M')<0$, we have 
$\uu-\uu'=\Omega \vec{e}_{z} \times ( \vec{M'}\vec{M} - \vec{O}_{+}\vec{M} - \vec{O}_{-}\vec{M})$, 
which leads to:
     \BE
	\vec{M'}\vec{M} \times (\uu-\uu') \vert_{n}=|\vec{M'}\vec{M}|^{2} \Omega 
	                  - \left( \vec{M'}\vec{M} \cdot (\vec{O}_{+}\vec{M} + \vec{O}_{-}\vec{M}) \right) \Omega  \,,
     \EE
giving (using $\vec{O}_{+}\vec{M} = \vec{O}_{+}\vec{O}_{-} + \vec{O}_{-}\vec{M}$):
    \BA     \label{eq:Eq-App-solGauss-tworot-12}
	\gth (M(\xx)) &=& - \frac{\Omega \Bn}{2\pi} 
	                 \int_{M' \ \textrm{in} \ P_{-}} \Bn' 
	                     \left( 1 - \frac{ \left( \vec{O}_{+}\vec{O}_{-} + 2 \ \vec{O}_{-}\vec{M} \right) \cdot\vec{M'}\vec{M}}
	                          {|\vec{M'}\vec{M}|^{2}}\right) \rmd \surf'  \nonumber \\ 
	   \nonumber \\
                                &=& \frac{\Omega \Bn}{2\pi} 
	                 \left( \Phi_{-}  +  \left( \vec{O}_{+}\vec{O}_{-} + 2 \ \vec{O}_{-}\vec{M} \right) \cdot \int_{M' \ \textrm{in} \ P_{-}} \Bn' 
	                      \frac{\vec{M'}\vec{M}}
	                          {|\vec{M'}\vec{M}|^{2}} \rmd \surf'  \right) \nonumber \\
	  \nonumber \\
                                &=&  \frac{\Omega \Bn}{2\pi} 
	                 \left( \Phi_{-}  -  \frac{ \left( \vec{O}_{+}\vec{O}_{-} + 2 \ \vec{O}_{-}\vec{M} \right) \cdot  \vec{O}_{-}\vec{M}}   {|\vec{O}_{-}\vec{M}|^{2}}  \ \Phi_{-}   \right) \nonumber \\
	  \nonumber \\                
			    &=& - \frac{\Omega \Bn \Phi_{-} }{2\pi} 
	   \left( 1 + \frac{\vec{O}_{-}\vec{M} \cdot \vec{O}_{+}\vec{O}_{-}}
	                   {|\vec{O}_{-}\vec{M}|^{2}}  \right)    \,,   
    \EA
for $\left(\Bn (M)>0, \Bn (M')<0 \right)$.\newline
The total helicity flux density in the positive polarity is obtained by summing 
\eq{Eq-App-solGauss-tworot-11} and \eq{Eq-App-solGauss-tworot-12} 
and supposing $\Phi_{+}=\Phi_{-}=\Phi_{0}$ to simplify:
    \BE
	\gth (M(\xx))= - \frac{\Omega \Bn }{2\pi}\ \frac{\vec{O}_{-}\vec{M} \cdot \vec{O}_{+}\vec{O}_{-}}{|\vec{O}_{-}\vec{M}|^{2}}\ \Phi_{0}   \,,
	                     \ \textrm{for} \ \Bn (M)>0 \,.
    \EE
	
Following the same derivation as above for $\Bn (M)<0$, we find \eq{Eq-Gth-tworot}.

%%% BIBLIOGRAPHY %%%%%%%%%%%%%%%%%%%%%%%%%%%%%%%%%%%%%%%%%%%%%%%%%%%%%%%%%%%
\	
	    % format of references provided by the journal (.bst)
\bibliographystyle{spr-mp-sola}
%\bibliographystyle{spr-mp-sola-cnd} %% Alternative style: no title,
                                      % no concluding page. 

     % name your Bibtex file containing your references (.bib)
\bibliography{PhotInj_dalmasse_etal}  

     % Checking: look if the file containing the ``\bibitem'' exits
     %           so check if the .bbl file exist (bibTeX compilation)
\IfFileExists{\jobname.bbl}{} {\typeout{}
\typeout{****************************************************}
\typeout{****************************************************}
\typeout{** Please run "bibtex \jobname" to obtain} \typeout{**
the bibliography and then re-run LaTeX} \typeout{** twice to fix
the references !}
\typeout{****************************************************}
\typeout{****************************************************}
\typeout{}}

\end{article} 

\end{document}